\documentclass[a4paper,journal]{IEEEtran}
\usepackage[pdftex]{color}
\usepackage[pdftex,final]{graphicx}
\usepackage[pdftex]{hyperref}
\usepackage{times}
\usepackage{amsfonts}
\usepackage{amssymb}
\usepackage{amsmath}
\usepackage{multirow}
\usepackage{rotating}
\newtheorem{thm}{Theorem}[section]
\newtheorem{theorem}{Theorem}[section]

\newtheorem{lemma}[thm]{Lemma}

\newtheorem{proposition}[thm]{Proposition}
\newtheorem{example}[thm]{Example}

\numberwithin{equation}{section}
\newcommand{\refeq}[1]{(\ref{eq:#1})}

\title{Capacity-achieving CPM schemes}

\author{Alberto~Perotti,~\IEEEmembership{Member,~IEEE,}
Alberto~Tarable,~\IEEEmembership{Member,~IEEE,} \\
Sergio~Benedetto,~\IEEEmembership{Fellow,~IEEE,} and
Guido~Montorsi,~\IEEEmembership{Senior~Member,~IEEE,} \\
Politecnico di Torino, Dipartimento di Elettronica\\
Corso Duca degli Abruzzi, 24 \\
I-10129 - Torino (Italy) \\
E-mail: \texttt{\{alberto.perotti, alberto.tarable, \\ sergio.benedetto, guido.montorsi\}@polito.it}}%

\begin{document}

\pagestyle{headings}

\renewcommand{\leftmark}{DRAFT: \today}

\maketitle

\begin{abstract}
The pragmatic approach to coded continuous-phase modulation (CPM) is proposed
as a capacity-achieving low-complexity alternative to the serially-concatenated
CPM (SC-CPM) coding scheme.
In this paper, we first perform a selection of the best spectrally-efficient
CPM modulations to be embedded into SC-CPM schemes. Then, we consider the pragmatic
capacity (a.k.a. BICM capacity) of CPM modulations and optimize it through a careful
design of the mapping between input bits and CPM waveforms.
The so obtained schemes are cascaded with an outer serially-concatenated convolutional
code to form a pragmatic coded-modulation system.
The resulting schemes exhibit performance very close to the CPM capacity without
requiring iterations between the outer decoder and the CPM demodulator. As a result,
the receiver exhibits reduced complexity and increased flexibility due to the separation
of the demodulation and decoding functions.
\end{abstract}

\section{Introduction}

Continuous phase modulation (CPM) is a class of bandwidth efficient
modulation schemes~\cite{bib:DigitalPhaseModulation} whose characteristic
of constant envelope makes them robust with respect to the nonlinearities
introduced by the analog baseband and radio-frequency sections of low-cost
transceivers usually found in consumer-type communications equipment.
Good application examples are the second generation GSM cellular system,
and satellite communication systems.

A CPM modulator is a finite-state machine delivering to the channel
a continuous-phase, constant envelope waveform that depends on its input
symbol and internal state.
In~\cite{bib:RimoldiDecomposition} a CPM modulator has been shown to be
decomposable into the cascade of a time-invariant convolutional encoder
(continuous-phase encoder, CPE) operating on a ring of integers, and of
a time-invariant memoryless modulator (MM).
Recently, this decomposition has been exploited by inserting an outer
convolutional encoder whose coded bits enter an interleaver
and then the CPE, thus forming what is known in the literature
as a serially-concatenated convolutional encoder (SCCC)~\cite{bib:Benedetto-Serial}.
In the following, we will call this scheme SC-CPM.

Iterating between the outer encoder and the CPE through the interleaver yields rather good
performance~\cite{bib:MoqvistSCCPM, bib:StuberNarayananTrellisCPM, bib:NarayananMSK},
which should be compared with the capacity of the CPM scheme.
Recently, the authors of~\cite{bib:LoeligerInfoRateMem}
proposed a simulation-based method for the computation of the capacity of channels with memory,
which can be applied to the evaluation of the capacity of the CPM schemes.   A similar
method has been proposed in ~\cite{bib:GeneralCPM}, where it has been applied to
a generalized form of CPM modulation achieving improved spectral efficiency.

In communication systems where the channel conditions can vary significantly with time,
an efficient radio resource management requires the availability at the physical layer of
adaptive coding-modulation, capable of varying its characteristics of bandwidth and
energy efficiency following the channel rate of variation.
This requirement has originated an active research on bit-interleaved (also known as
\emph{pragmatic}) coded modulation \cite{bib:ZehaviPragmatic, bib:ZehaviTrellis, bib:BICM}.
Pragmatic coded modulation consists in cascading a highly performing versatile binary encoder
(typically, a punctured turbo or low-density parity-check code)
with several modulation schemes with increasingly large signal alphabets.
The versatile encoder is capable of varying both its rate and codeword length in a wide range
in order to achieve increasing spectral efficiencies.
An example of the obtainable results has been published in \cite{bib:MHOMS}, which demonstrates
a scheme based on SCCC and linear two-dimensional modulations yielding spectral efficiencies
in a very wide range lying around 1 dB from the Shannon capacity limits.
This paper deals with the extension of the pragmatic approach to CPM modulation (called P-CPM in the following).
This approach does not require iterations between the outer encoder and the
CPE, since the CPM is treated exactly as a linear modulation in a bit-interleaved turbo-trellis
coded modulation.
A nice consequence is that the overall CPE state complexity is not enhanced by the number of
iterations, thus permitting to increase the bandwidth efficiency through the use of a larger
number of CPE states.

In~\cite{bib:BenedettoITA2007} a first attempt to design pragmatic schemes employing
CPM modulation has been presented.
The authors showed that the pragmatic capacity of CPM schemes heavily depends on the
mapping between information bits and CPM signals, and presented a simulation-based mapping
optimization algorithm with some examples.
Further improvements have been presented in~\cite{bib:BenedettoGlobecom2007}, where
a procedure for optimizing such mapping has been proposed.  The optimized CPM modulators
have been embedded into a P-CPM scheme and its performance has been assessed through simulation.
For the considered CPM schemes, performance improvements of more than 2 dB in pragmatic
capacity have been observed.

Previuos literature on the subject includes~\cite{bib:ShaneWeselPCCC-CPM}, where a scheme
consisting of a parallel concatenated turbo code and continuous-phase modulation has been
investigated and a modified encoder has been proposed.
In~\cite{bib:giann05}, CPM modulations have been studied in the
context of multiple antenna systems with layered space-time coding. Reduced-complexity receivers
have been proposed and differential encoding has been introduced in order to obtain an
increased coding gain.

This paper presents a systematic and comprehensive approach to the problem of designing capacity
approaching SC-CPM and P-CPM, through the evaluation of the CPM and P-CPM capacities, the search
for optimal (in the sense of offering the best trade-off between bandwidth efficiency, energy
efficiency and complexity) CPM modulations, their embedding into SC-CPM and P-CPM schemes through the optimization of the mapping between coded bits and CPM waveforms, and a thorough comparison of their error probability performance with respect to capacities. Examples refer to rectangular and raised cosine frequency waveforms, and to three spectral efficiencies deemed important for the applications.

This paper is organized as follows: in Sec.~\ref{sec:system-description} a general
description of the considered coding and modulation systems is given.
Sec.~\ref{sec:compcap} presents the procedure used for the CPM capacity computation
through simulation. In Sec.~\ref{sec:opt} the optimization procedure used to select
the best CPM parameters to be used in SC-CPM schemes is defined.
Sec.~\ref{sec:pragcapopt} presents the procedure used for the optimization of P-CPM
schemes: first, the optimal mapping is derived, the corresponding optimized CPE scheme
is given and then the CPM schemes with best pragmatic capacity are selected.
Sec.~\ref{sec:implementation} shows how the optimized CPM schemes have been embedded
into SC-CPM and P-CPM coded systems.
Finally, Sec.~\ref{sec:results} shows the results obtained using the selected CPM
modulations in SC-CPM and P-CPM schemes.

%%%%%%%%%%%%%%%%%%%%%%%%%%%%%%%%%%%%%%%%%%%%%%%%%%%%%%%%%%%%%%%%%%%%%%
\section{System description}
\label{sec:system-description}
%%%%%%%%%%%%%%%%%%%%%%%%%%%%%%%%%%%%%%%%%%%%%%%%%%%%%%%%%%%%%%%%%%%%%%

A CPM modulator is a device with memory that generates continuous-phase,
constant-envelope modulated waveforms

\begin{equation}
x(t) = \sqrt{\frac{2E_s}{T}} e^{j \psi(t)}
\end{equation}

\noindent whose phase

\begin{equation}
\psi(t) = 2 \pi h \sum_{n = -\infty}^\infty{a_n q(t - nT)}
\end{equation}

\noindent depends on the input information symbols $a_n \in \{\pm 1,
\pm 3, \ldots, \pm (M - 1) \}$, where $M = 2^m$ is the size of the
input alphabet. Here, $T$ is the symbol interval, $E_s$ is the energy
per symbol, $h = Q/P$ is the \emph{modulation index} ($Q$ and $P$ are
relatively prime integers), and $q(t)$ is the phase pulse, a
continuous function with the following properties

\[q(t) = \left\{\begin{tabular}{ll}
0           &   $t \leq 0$ \\
$\frac{1}{2}$ & $t \geq LT$
\end{tabular}
\right.\]

\noindent The phase pulse is usually defined as the integral of a
\emph{frequency pulse} $s(t)$

\[q(t) = \int_{-\infty}^t{s(\tau) d \tau}\]

\noindent A CPM scheme is then defined by specifying its
parameters $M$, $h$, $L$ and the frequency pulse $s(t)$.

In this paper, we will consider rectangular (REC) and raised-cosine
(RC) frequency pulses. The REC pulse is defined as

\[s_{\rm REC}(t)=\frac{1}{2LT} [u(t) - u(t - LT)] \]

\noindent and the RC pulse is defined as

\[s_{\rm RC}(t)=\frac{\pi}{4LT} [u(t) - u(t - LT)]
\left[1 - \cos\left(\frac{2 \pi t}{LT}\right)\right] \]

\noindent where $u(t)$ is the unitary step function.

\begin{figure}[!ht]
\centering
\includegraphics[width=3.5in]{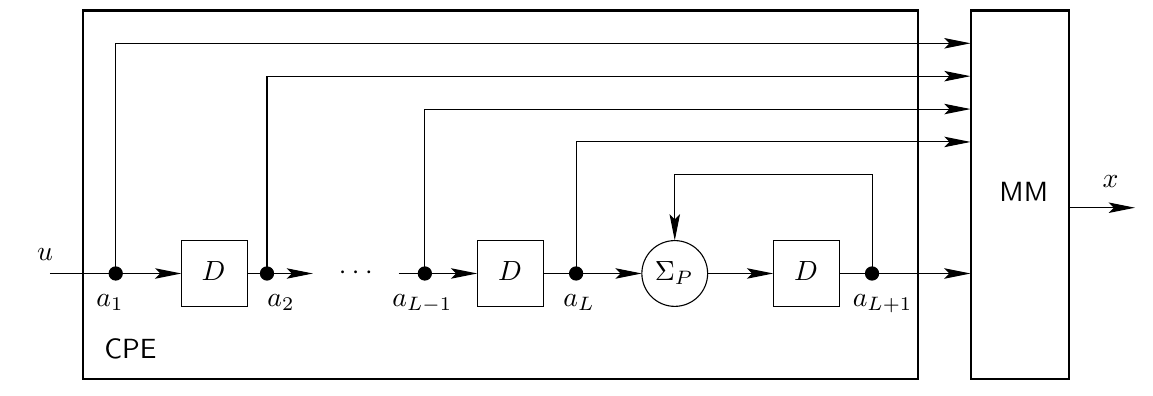}
\caption{Rimoldi decomposition of the CPM encoder. The block labelled
$\Sigma_P$ is a modulo-$P$ adder.}
\label{fig:cpmenc}
\end{figure}

According to the well known Rimoldi decomposition
\cite{bib:RimoldiDecomposition}, the modulator can be represented as
the cascade of a continuous phase encoder (CPE) and a memoryless
modulator (MM) as in Fig.~\ref{fig:cpmenc}. The CPE, in general, is a time-invariant convolutional
encoder operating on a ring of integers.

\begin{figure}
\centering
\includegraphics[clip,width=3.5in]{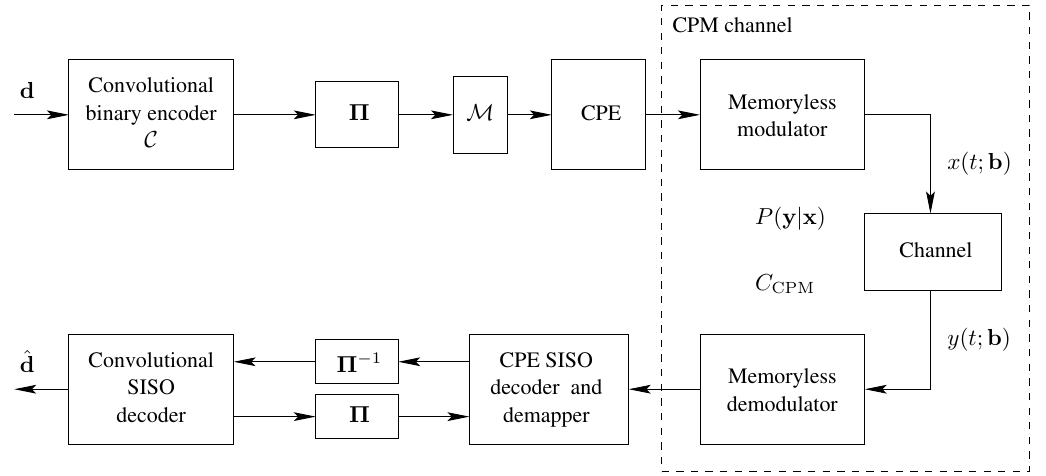}
\caption{Block diagram of a serially concatenated CPM co-decoder.
The block labeled $\cal M$ maps blocks of $m$ bits to $M$-ary CPM symbols.}
\label{fig:SCCPMcodec}
\end{figure}

The serial structure of the Rimoldi decomposition has been exploited by adding an
outer convolutional encoder connected to the CPE trough an interleaver, so as to form a serially concatenated convolutional encoder (SCCC) with interleaver \cite{bib:ett} (see Fig.~\ref{fig:SCCPMcodec}).  The iterative receiver performs decoding iterations
between the inner SISO decoder, which operates on the CPE trellis, and
the outer SISO decoder, which operates on the outer convolutional code
trellis. This way,  very good performance can be achieved, to be compared with the capacity of the CPM scheme. In the following, we will denote this scheme as serially concatenated CPM (SC-CPM)~\cite{bib:MoqvistSCCPM}.

\begin{figure}
\centering
\includegraphics[clip,width=3.5in]{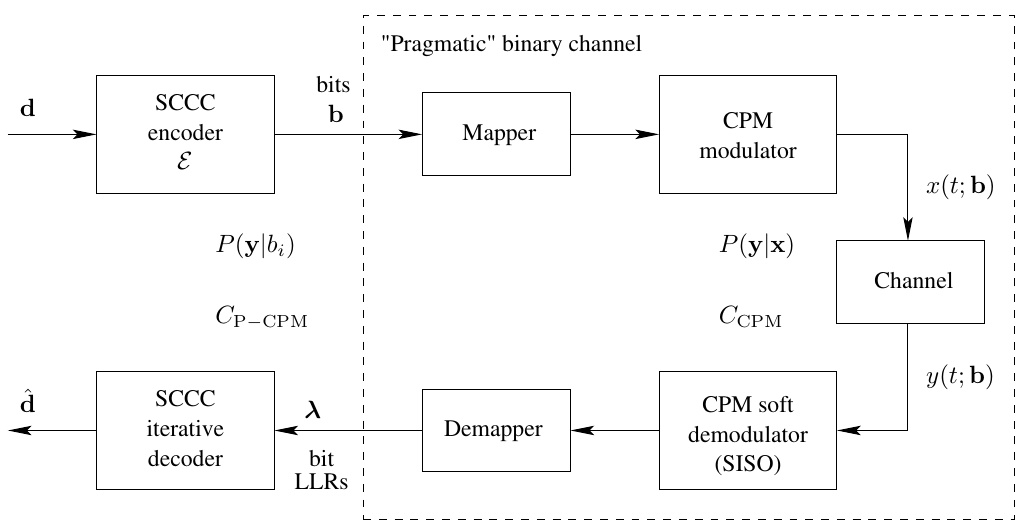}
\caption{Block diagram of an encoded CPM mo-demodulator used with
a pragmatic approach.} \label{fig:CPMcodec}
\end{figure}

In this paper, we  also consider  a \emph{pragmatic} approach, which results in a coded
modulation scheme equivalent to a \emph{bit-interleaved} coded
modulation (BICM, see Fig.~\ref{fig:CPMcodec}).    It consists in
cascading a powerful variable-rate binary encoder (a serially-concatenated convolutional code (SCCC) with interleaver  in this paper) with
the CPM modulator. The CPM modulator input is connected to the output
of the channel encoder  that computes the binary sequence
$\mathbf{b} = (b_i, i \in \mathbb{Z})$ from the binary information
sequence $\mathbf{d}$. Then,the coded binary sequence is mapped to the CPM modulator
that generates a corresponding CPM signal sequence $x(t; \mathbf{b})$. We will call this scheme pragmatic CPM (P-CPM).

The channel is an additive white Gaussian noise (AWGN) channel, whose output signal is $y(t; \mathbf{b}) = x(t; \mathbf{b}) + n(t)$, where $n(t)$ is a zero-mean white Gaussian process with two-sided power spectral density $N_0 / 2$.   At the receiver, the CPM soft demodulator and demapper provide the sequence
{\boldmath $\lambda$}~=~$(\lambda_i, i \in \mathbb{Z})$ of
log-likelihood ratios on the information bit sequence to the outer
decoder

\[\lambda_i = \log\frac{P(b_i = 1 | y(t; \mathbf{b}))}{P(b_i = 0 |
y(t; \mathbf{b}))}\]

\noindent Finally, {\boldmath $\lambda$} is used by the outer iterative decoder
to compute the information sequence estimate
$\mathbf{\hat{d}}$.

The main difference between P-CPM and SC-CPM is that no iterations are performed
between the inner CPM demodulator and the outer turbo decoder. Thus, it keeps the nice features of all pragmatic approaches, which merge independent binary codes with higher order modulations without requiring joint optimization.

Our goal is to design coded modulation schemes based on CPM capable of transmitting
information at rates close to the CPM channel
capacity\footnote{Hereafter, with \emph{CPM channel capacity} we mean
\emph{constrained} capacity with uniform input distribution.}.
Such goal will be pursued by analyzing the P-CPM and SC-CPM schemes
and choosing the best trade-off between the following characteristics:

\begin{itemize}
 \item Spectral efficiency.
 \item Energy efficiency.
 \item Decoding complexity.
\end{itemize}

Spectral efficiency and energy efficiency are two well-known concepts.
The decoding complexity is defined here as the overall number of trellis
edges per information bit visited by the decoding algorithm. This definition depends on the type of
coded modulation being used (i.e., P-CPM or SC-CPM).   For the
P-CPM scheme, we have

\begin{equation}
\label{eq:complexity-pcpm}
\mathcal{Y}_{\rm P-CPM} = \mathcal{Y}_{\rm SCCC} + \mathcal{Y}_{\rm
CPM}
\end{equation}

\noindent where

\[\mathcal{Y}_{\rm CPM} = M N_{s,CPE} \frac{N_o}{m K_o}\]

\noindent and $N_o$  and $K_o$ are, respectively, the outer code
word length and information word length, and $N_{s,CPE} = P
M^{L-1}$ is the number of states of the CPE. The complexity of the
SCCC binary decoder is  $\mathcal{Y}_{\rm SCCC}$ and will be
defined later according to the chosen code (see (\ref{eq:SCCCComplexity})).

For the SC-CPM scheme, we adopt the following definition, which holds
for binary convolutional codes with rate $\frac{K_o}{N_o}$ obtained by
puncturing a rate 1/2 mother code:

\begin{equation}
\label{eq:complexity-sccpm}
\mathcal{Y}_{\rm SC-CPM} = N_{it}\left(2 N_{so} + \mathcal{Y}_{\rm
CPM}\right)
\end{equation}

\noindent where $N_{it}$ is the number of decoding iterations between the
outer SISO and the inner CPE SISO, and $N_{so}$ is the number of states of
the outer convolutional encoder.

%
% Both in the pragmatic scheme and in the SC-CPM scheme, the
%theoretical
% limit to the quantity of information that can be transferred is the
% capacity of the CPM scheme. In the following, we will assume that
%the
% input sequences to the CPM modulator are equally likely, hence we
%are
% interested in the \emph{mutual information} between $\mathbf{b}$ and
% {\boldmath $\lambda$} with such assumption.

In the following, we will optimize the CPM schemes to be embedded into
both SC-CPM and P-CPM starting from their capacity evaluation.

\section{CPM capacity computation}
\label{sec:compcap}

With reference to Figures~\ref{fig:SCCPMcodec} and \ref{fig:CPMcodec}, CPM signals are
infinite-length waveforms.  Consider then a finite observation window
$[-NT,NT]$ of $2N+1$ symbol intervals, and define the channel mutual
information over it:
\[
I({\rm \bf X};{\rm \bf Y})=E_{\mathbf{x}, \mathbf{y}
}\left\{\log_2\frac{p_{\bf Y|X}({\rm \bf y}|{\rm \bf x})}{p_{\bf
Y}({\rm \bf y})}\right\}
\]

\noindent where ${\rm \bf X}$ is a vector of  samples of $x(t)$
that form a sufficient statistic of $x(t)$ in the  interval
$[-NT,NT]$ and ${\rm \bf Y}$ is the correspondent vector of
channel outputs. The set of all possible values for ${\rm \bf X}$
will be denoted $\mathcal X$ throughout the paper. The CPM
capacity can be defined through the limit

\begin{equation}
C_{\mathrm{CPM}}=\lim_{N\rightarrow\infty}\frac{1}{(2N+1)T}I({\rm \bf
X};{\rm \bf Y}) \;\;{\rm [bits/s]}\label{eq:a}
\end{equation}

Using the definition of mutual information $I$ and of the log-likelihood ratio (LLR) $$\lambda({\rm \bf
u},{\rm \bf y} )\triangleq \log_2(p_{\bf Y|X}({{\rm  \bf y}|{\rm \bf
u}))-\log_2(p_{\bf Y|X}({\rm \bf y}|{\rm \bf u}_{\rm ref}))}$$

\noindent as well as the $\max^*$ operator~\cite{bib:ett}

    \[ {\max}^*(a,b) \triangleq  \log_2(2^a+2^b)
    %\footnote{For notational simplicity here the
    %base of the logarithm in the definition of $\max^*$ is 2 instead
    %of $e$.}
    \]
we can transform \refeq{a} into
    \[
    C_{\mathrm{CPM}} =m - \lim_{N\rightarrow\infty}
    \frac{1}{(2N+1)T} E_{\mathbf{x}, \mathbf{y}}\left\{{\max_{\rm \bf u \in \mathcal{X}}}^*
    \lambda({\rm \bf u},{\rm \bf y})-\lambda({\rm \bf x},{\rm \bf y})\right\}.
    \]
where $m$ is the number of bits per CPM symbol.

It has been observed in \cite{bib:LoeligerInfoRateMem} that the
first term inside the average is a by-product of the SISO
algorithm and corresponds to the $\max^*$ of the forward path
metrics $\alpha$ at step $N$ in the CPE decoder
\[
 {\max_{\rm \bf u\in \mathcal{X}}}^* \lambda({\rm \bf u},{\rm \bf
y})={\max_{s}}^*\alpha_N(s),
\]
with the following initialization

\[ \alpha_{-N}(s)=\left\{
    \begin{array}{ll}
        0, & s=0; \\
        -\infty, & s\neq 0.
    \end{array}
\right.
\]
where $s$ runs in the set of trellis states of the CPE.
\noindent Moreover, since the channel is  memoryless, the second term
is
obtained by summing the LLRs of the transmitted waveforms:
\[
\lambda({\rm \bf x},{\rm \bf y})=\sum_{i=-N}^N \lambda(x_i,y_i).
\]

Finally, invoking the ergodic properties of the system, the
ensemble average can be removed leading to:
\[
C_{\mathrm{CPM}} =m - \lim_{N\rightarrow\infty}
  \frac{1}{2N+1} \left({\max_{\bf u}\in \mathcal{X}}^*
  \lambda({\rm \bf u},{\rm \bf y})-\lambda({\rm \bf x},{\rm \bf y})\right).
\]
Thus, the CPM capacity can be estimated through a Monte-Carlo
simulation of the three internal blocks of
Fig.~\ref{fig:CPMcodec}, i.e., the CPM modulator, the channel,
 and the CPM soft demodulator, followed by
a time average.

\section{Optimization procedure}
\label{sec:opt}

We define the complexity of a CPM scheme as the number of edges per
CPM input bit\footnote{Please note that the definition of
$\mathcal{Y}$ differs from $\mathcal{Y}_{\rm CPM}$ in that the former
is the number of CPE trellis edges per CPM input bit, while the latter
is the number of CPE trellis edges per bit at the channel encoder
input.} ${\mathcal Y} = P 2^{mL} / m$,
where the parameters have been defined in Section II. Our aim is to
maximize the CPM capacity measured in bits/s/Hertz versus the
signal-to-noise ratio for a given complexity.   The optimization
algorithm modifies the CPM
parameters ($m$, $P$, and $L$) yielding the given complexity, and
computes the CPM capacity expressed in bits per CPM waveform for a
given signal-to-noise ratio according to the algorithm described in Sec.~\ref{sec:compcap}.
Then, based on a bandwidth definition
of the CPM scheme, it evaluates the rate and chooses the best scheme
for all signal-to-noise ratios of interest.  In the following, we
describe  the optimization algorithm step by step:

\begin{enumerate}
    \item Define the bandwidth $B$ of the CPM system that contains a
given percentage of the total signal power.
    \item Define the CPM symbol rate as $R_s\triangleq 1/T$.
        \item The {\em symbol} signal-to-noise ratio $E_s/N_0$ is
    given by
    \[
    \frac{E_s}{N_0}=\frac{1}{R_s}\frac{P_T}{N_0}
    \] where $P_T$ is the transmitted power.
    \item Compute the capacity (in bits/symbol) of each CPM scheme using the method described in
Section~\ref{sec:compcap} as a function of the
     symbol SNR $C_{\mathrm{CPM}}=C_{\mathrm{CPM}}(E_s/N_0)$.
    \item Evaluate the CPM capacity as
    \begin{equation}\label{eq:Cdef}
        C = R_s
C_{\mathrm{CPM}}\left(\frac{1}{R_s}\frac{P_T}{N_0}\right)\;\;\text{[bits/s/Hz]
}.
    \end{equation}
       \item All CPM schemes with a given complexity ${\mathcal Y}$  are compared
with respect to $C$ for each value of $\frac{P_T}{N_0}$ and the
best is chosen.
        \item Finally, to obtain the normalized plot that
uses the {\em bit} SNR $E_b/N_0$ on the abscissa we use the
relationship
    \begin{equation}\label{eq:ebn0def}
\frac{E_b}{N_0} = \frac{1}{C_{\mathrm{CPM}}}\frac{E_s}{N_0}=\frac{R_s}{C}
\frac{E_s}{N_0} = \frac{1}{C} \frac{P_T}{N_0}.
    \end{equation}
\end{enumerate}

\begin{figure}[ht]
\centering\includegraphics[angle=-90,width=3.5in]{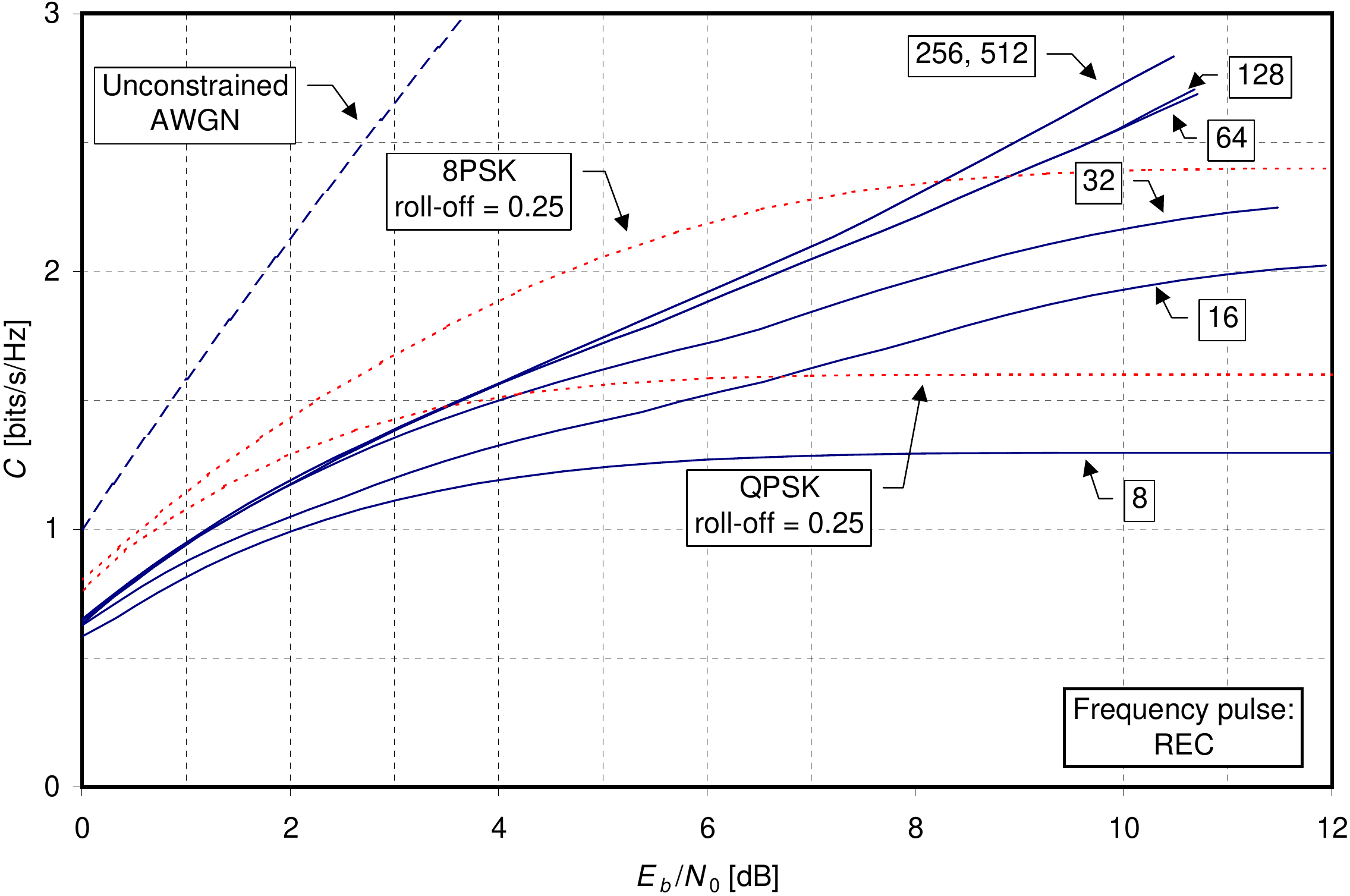}
\caption{Capacity versus $E_b/N_0$ of the best CPM schemes
with rectangular frequency pulse and variable complexity. The
bandwidth is defined at 99 \% of the total power.}
\label{fig:SpecEffREC}
\end{figure}

\begin{figure}[ht]
\centering\includegraphics[angle=-90,width=3.5in]{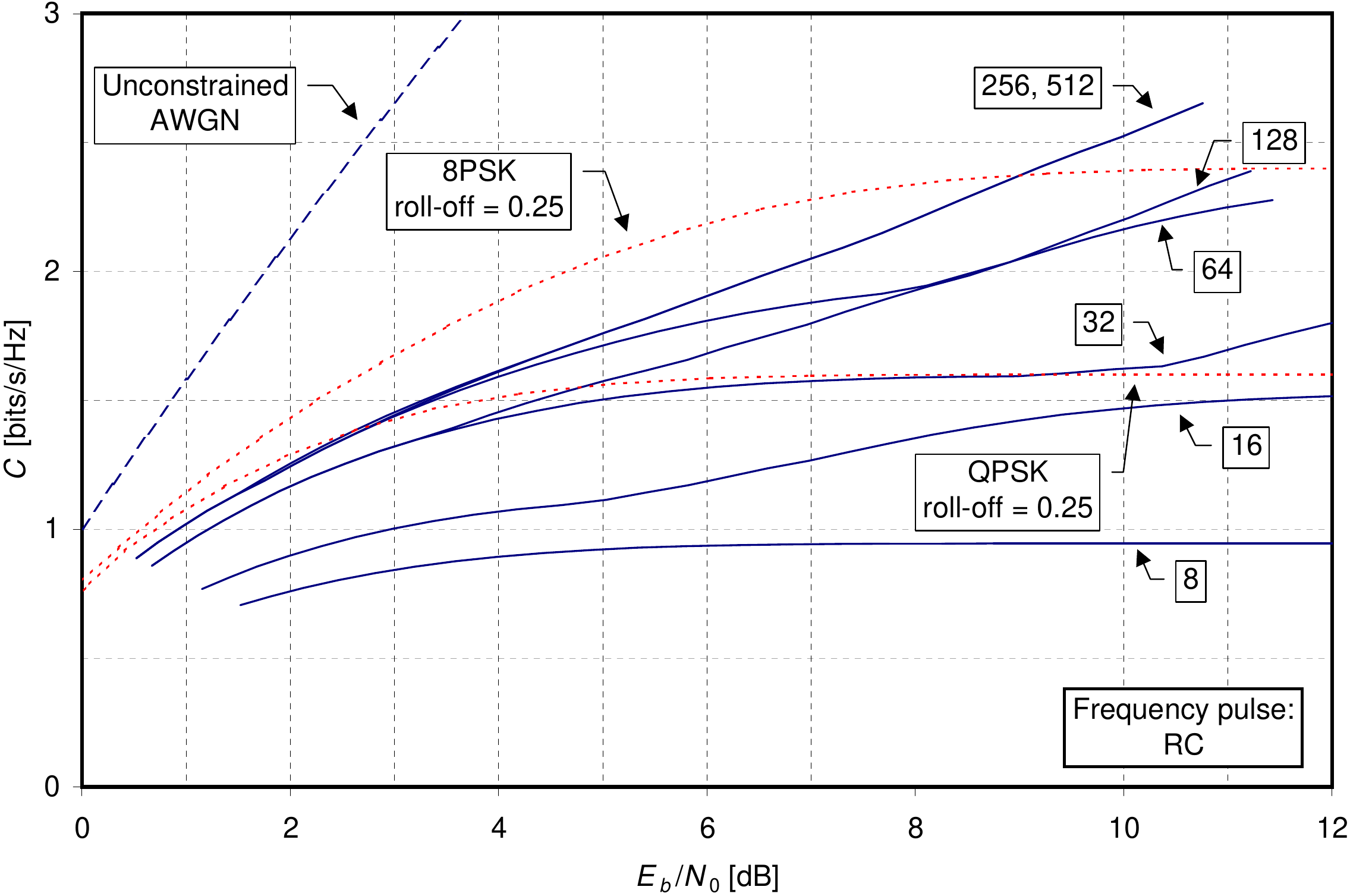}
\caption{Capacity versus $E_b/N_0$ of the best CPM
schemes with raised-cosine frequency pulse and variable complexity.
The bandwidth is defined at 99 \% of the total power.}
\label{fig:SpecEffRC}
\end{figure}

In Fig.~\ref{fig:SpecEffREC} and Fig.~\ref{fig:SpecEffRC}  we show the
capacity versus $E_b/N_0$ for the best CPM schemes with REC and RC frequency
pulses and complexity ranging from 8 to 512. The bandwidth is defined as the one including
99\% of the total signal power. For comparison purposes, in the figure we have
also plotted the capacity of QPSK and 8PSK modulations
with a square root raised cosine shaping filter with roll-off 0.25, and the unconstrained
capacity of the AWGN channel. 

In Table~\ref{tab:bestrec99} and Table~\ref{tab:bestrc99} we list for
each SNR  ($E_s/N_0$) and complexity from 8 to 512 the CPM schemes with
REC and RC frequency pulses
achieving the highest capacity for a given signal-to-noise ratio.
Each scheme is characterized by the parameters $E_b/N_0$ as defined in (\ref{eq:ebn0def}),
the capacity $C$ as defined in (\ref{eq:Cdef}), the CPM parameters $m$, $L$, $P$, and finally the
symbol rate $R_s$.

\begin{table*}[ht]
\centering\includegraphics[angle=-90,width=7in]{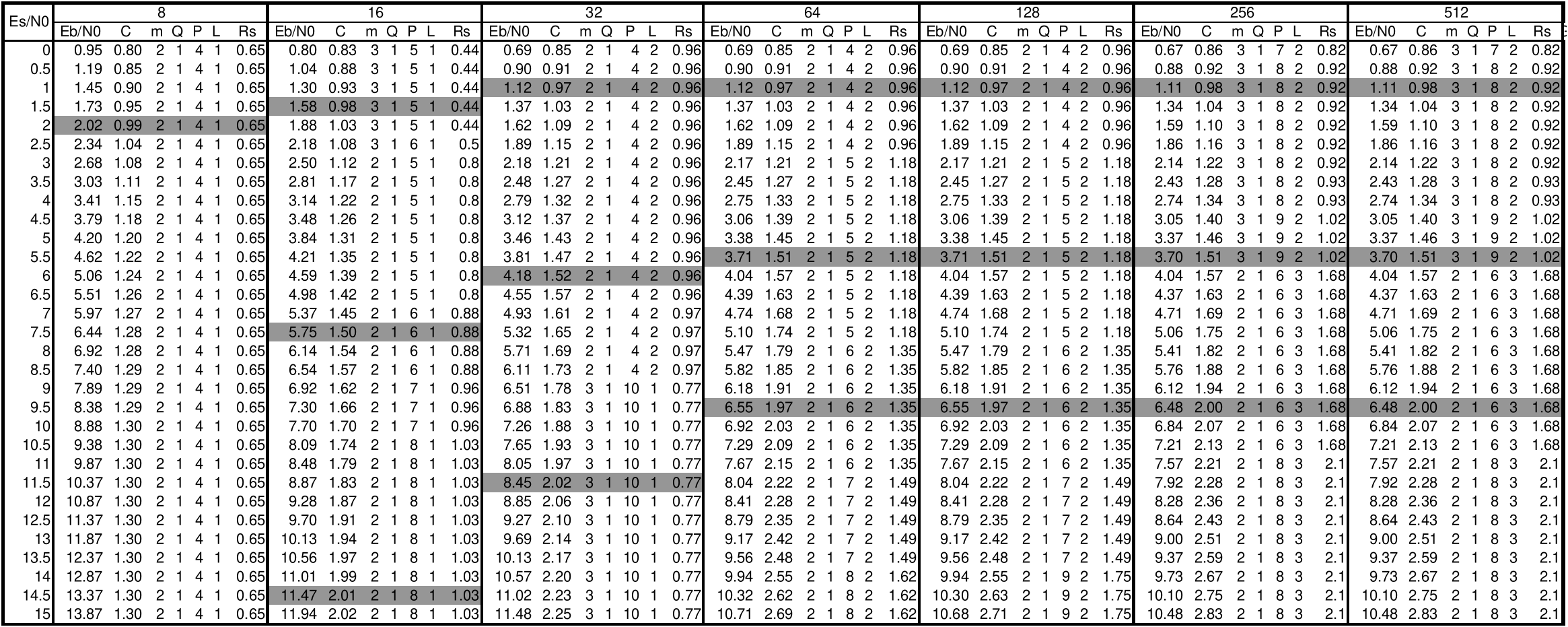}
\caption{Table of the CPM schemes with best $C$ and rectangular
frequency pulse at 99\% bandwidth. The complexity range is 8 to 512. The highlighted entries
correspond to the selected CPM schemes for target capacities of 1, 1.5 and 2 bits/s/Hz.}
\label{tab:bestrec99}
\end{table*}

\begin{table*}[ht]
\centering\includegraphics[angle=-90,width=7in]{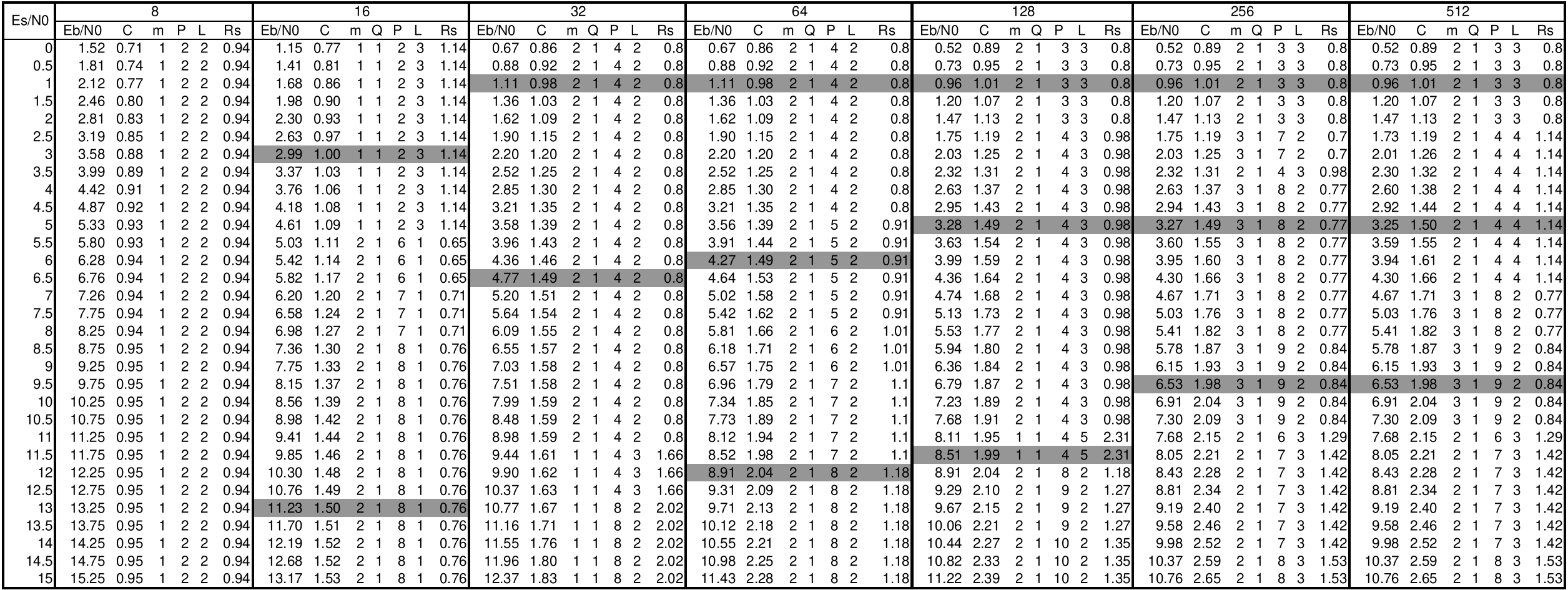}
\caption{Table of the CPM schemes with best $C$ and raised-cosine
frequency pulse at 99\% bandwidth. The complexity range is 8 to 512. The highlighted entries
correspond to the selected CPM schemes for target capacities of 1, 1.5 and 2 bits/s/Hz.}
\label{tab:bestrc99}
\end{table*}

%%%%%%%%%%%%%%%%%%%%%%%%%%%%%%%%%%%%%%%%%%%%%%%%%%%%%%%%%%%%%%%%%%%%%%
\section{Pragmatic capacity of a TCM scheme}
\label{sec:pragcapopt}
%%%%%%%%%%%%%%%%%%%%%%%%%%%%%%%%%%%%%%%%%%%%%%%%%%%%%%%%%%%%%%%%%%%%%%

%In the previous section we have found the best CPM schemes for a
%given rate and signal-to-noise ratio. Here, we propose an
%algorithm that, starting form the capacity-optimized CPM scheme,
%finds the mapping that maximizes the pragmatic capacity.

With reference to Fig.~\ref{fig:CPMcodec}, let the binary sequence
{\bf b}  be formed by vectors ${\rm \bf B}_n = (B_{n,1}, \dots,
B_{n,m})$ of the $m$ bits entering the CPM modulator at the $n$-th
trellis step. As previously, we consider a length-$(2N+1)$ input
sequence of ${\rm \bf B}_n$.

The scheme of Fig.~\ref{fig:CPMcodec} can be characterized by a
\emph{pragmatic} capacity $C_{\mathrm{P-CPM}}$,
 defined by:
\begin{equation}
C_{\mathrm{P-CPM}} \triangleq
\lim_{N\rightarrow\infty}\frac{1}{2N+1} \sum_{n=-N}^N \sum_{i=1}^m
I(B_{n,i};{\rm \bf Y}), \label{eq:cpcpm}
\end{equation}
which depends on the \emph{mapping} between input binary sequences
and CPM waveforms.

\begin{figure}[!ht]
\centering
\includegraphics[angle=270,width=3.5in]{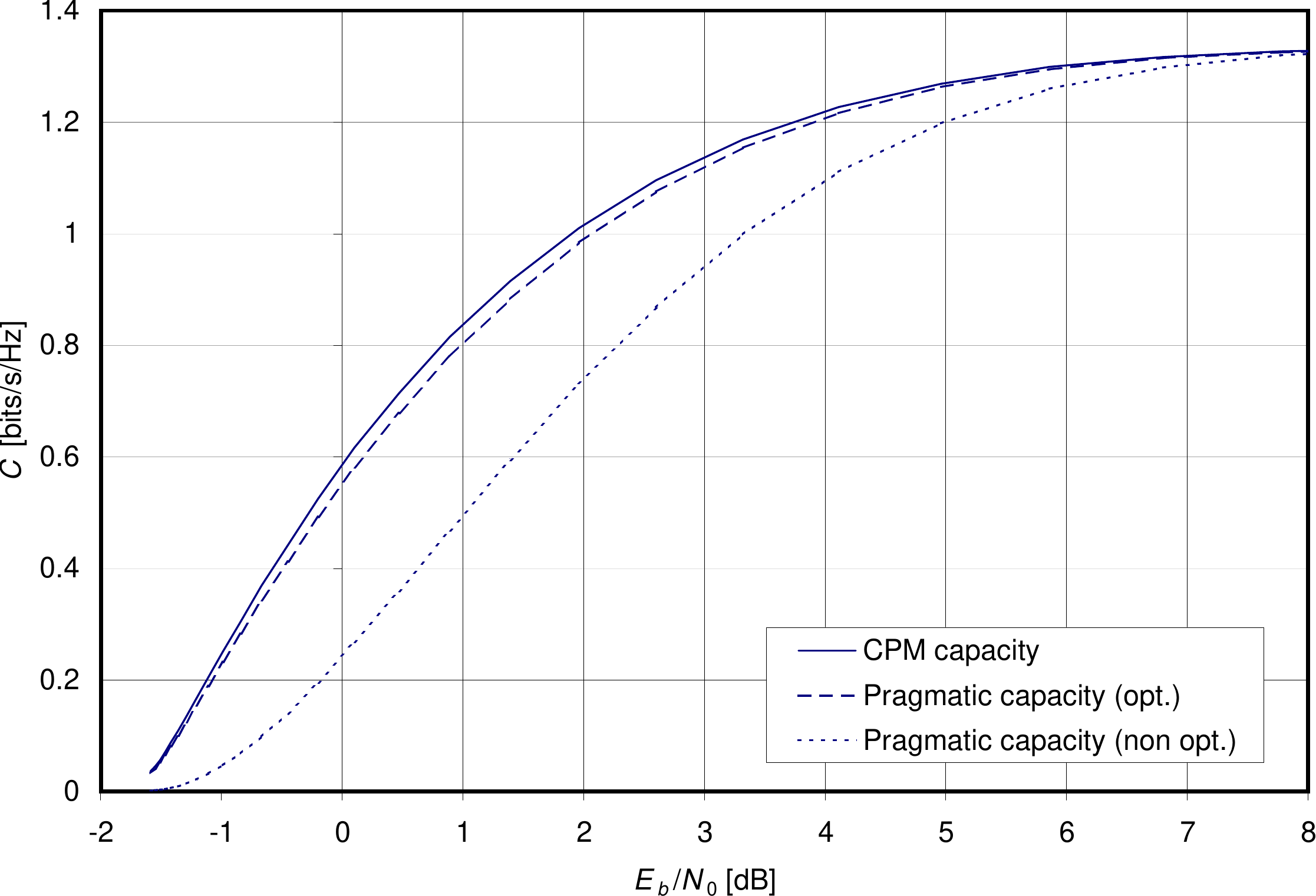}
\caption{CPM capacity and pragmatic capacities of the binary REC
scheme with $h = 1/2$ and $L = 3$.}
\label{fig:cap-comp}
\end{figure}

In order to optimize the system performance, it is crucial
to search for the optimal mapping, i.e., the one that maximizes
the pragmatic capacity of a given CPM scheme. To this purpose, we will derive in the following a lower
bound to the pragmatic capacity, which makes explicit the
dependence on the mapping. Before delving deeper into the analysis, though,
we show in Fig.~\ref{fig:cap-comp} a comparison between the optimized P-CPM capacity,
obtained through the algorithm derived in the following, and that obtained by a
straightforward application of the mapping induced by the Rimoldi decomposition.
The curves show that the pragmatic capacity of the optimized
CPE is significantly improved with respect to the non optimized CPE.
For the non optimized CPE, the gap between the CPM capacity and the
pragmatic capacity is roughly 1.5 dB, while the pragmatic capacity of
the optimized CPE is very close to the CPM capacity.
The obtained gain is roughly 1.4 dB.

Since
\begin{equation}
I(B_{n,i};{\rm \bf Y}) = 1 - H(B_{n,i}|{\rm \bf Y}), \label{eq:ih}
\end{equation}

\noindent the optimal mapping is the one that yields a minimum of $\sum_{n}
\sum_{i} H(B_{n,i}|{\rm \bf Y})$. Define $\mathcal{X}_{n,i}(b)$,
$b=0,1$, as the set of CPM waveforms whose mapping satisfies
$B_{n,i} = b$. Obviously, $\mathcal{X} = \mathcal{X}_{n,i}(0) \cup
\mathcal{X}_{n,i}(1)$. We have:

\begin{equation} \label{eq:hby}
H(B_{n,i}| \mathbf{Y} )  = \int H(B_{n,i}|\mathbf{Y}=\mathbf{y})
p_{\mathbf{Y}}(\mathbf{y}) d\mathbf{y}.
\end{equation}

Now, $H(B_{n,i}|\mathbf{Y}=\mathbf{y})$ is given explicitly by:

\begin{equation} \label{eq:hby1}
H(B_{n,i}|\mathbf{Y}= \mathbf{y}) = H\left( \frac{\sum_{\mathbf{u}
\in \mathcal{X}_{n,i}(0)} p_{\mathbf{Y} | \mathbf{X}}(\mathbf{y} |
\mathbf{u})}{\sum_{\mathbf{u} \in \mathcal{X}} p_{\mathbf{Y} |
\mathbf{X} }(\mathbf{y} | \mathbf{u})}\right),
\end{equation}

\noindent where the binary entropy function $H(p) =-p\log p -(1-p)\log(1-p)$
has been used in the RHS of the above equation.

To find a viable path to mapping optimization, we approximate the
distribution of $\mathbf{Y}$ with its value for infinite SNR, by
making in (\ref{eq:hby}) the following substitution:

\begin{equation}
p_{\mathbf{Y}}(\mathbf{y}) \approx \frac{1}{|\mathcal{X}|}
\sum_{\mathbf{x}
 \in \mathcal{X}} \delta(\mathbf{x} -\mathbf{y}), \label{eq:hsnr}
\end{equation}

\noindent where $|\mathcal{X}|$ is the size of $\mathcal{X}$ and
$\delta(\cdot)$ is Dirac's delta. The above approximation makes
sense because the optimal mapping for large SNR should reasonably
be the same as for infinite SNR.

Substituting \refeq{hsnr} into \refeq{hby}, we obtain:

\begin{equation}
\begin{split}
H(B_{n,i}|\mathbf{Y}) \approx & \frac1{2} \sum_{b=0,1}
\frac{2}{|\mathcal{X}|} \times \\ & \sum_{\mathbf{x} \in \mathcal{X}_{n,i}(b)}
H\left( \frac{\sum_{\mathbf{u} \in \mathcal{X}_{n,i}(0)} p_{\mathbf{Y}
| \mathbf{X}}(\mathbf{x} | \mathbf{u})}{\sum_{\mathbf{u} \in
\mathcal{X}} p_{\mathbf{Y} |
\mathbf{X} }(\mathbf{x} | \mathbf{u})}\right),
\end{split}
\end{equation}

\noindent where the sum w.r.t $\mathbf{x}$ over $\mathcal{X}$ has been split
into separate sums over $\mathcal{X}_{n,i}(0)$ and
$\mathcal{X}_{n,i}(1)$.

By Jensen's inequality and the concavity of the entropy function,
we obtain the upper bound:
\begin{equation} \label{eq:hb0y1}
H(B_{n,i}|\mathbf{Y}) \leq \frac1{2} \sum_{b=0,1} H\left(
\frac{2}{|\mathcal{X}|} \sum_{\mathbf{x}
 \in \mathcal{X}_{n,i}(b)}\frac{\sum_{\mathbf{u} \in
\mathcal{X}_{n,i}(0)} p_{\mathbf{Y} | \mathbf{X}}(\mathbf{x} |
\mathbf{u})}{\sum_{\mathbf{u} \in \mathcal{X}} p_{\mathbf{Y} |
\mathbf{X} }(\mathbf{x} | \mathbf{u})}\right).
\end{equation}

Owing to the ergodic property of the system, it can be shown that
$\sum_{\mathbf{u} \in \mathcal{X}} p_{\mathbf{Y} | \mathbf{X}
}(\mathbf{x} | \mathbf{u})$ is equal to a constant for almost every
transmitted $\mathbf{x}$. Thus:
\begin{equation} \label{eq:hb0y2}
H(B_{n,i}| \mathbf{Y} ) \leq  \frac1{2} \sum_{b=0,1} H\left(
\frac{2\sum_{\mathbf{x}
 \in \mathcal{X}_{n,i}(b)}\sum_{\mathbf{u} \in
\mathcal{X}_{n,i}(0)} p_{\mathbf{Y} | \mathbf{X}}(\mathbf{x} |
\mathbf{u})}{\sum_{\mathbf{x}
 \in \mathcal{X}}\sum_{\mathbf{u} \in
\mathcal{X}} p_{\mathbf{Y} | \mathbf{X}}(\mathbf{x} | \mathbf{u})}
\right).
\end{equation}

Finally, notice that, since $p_{\mathbf{Y} |
\mathbf{X}}(\mathbf{x} | \mathbf{u}) = p_{\mathbf{Y} |
\mathbf{X}}(\mathbf{u} | \mathbf{x})$ for the AWGN channel, the two
entropies in the RHS of \refeq{hb0y2} have the same arguments, and
thus are equal:
\begin{equation} \label{eq:hb0y3}
H(B_{n,i}| \mathbf{Y} ) \leq  H\left( \frac{2\sum_{\mathbf{x}
 \in \mathcal{X}_{n,i}(0)}\sum_{\mathbf{u} \in
\mathcal{X}_{n,i}(0)} p_{\mathbf{Y} | \mathbf{X}}(\mathbf{x} |
\mathbf{u})}{\sum_{\mathbf{x}
 \in \mathcal{X}}\sum_{\mathbf{u} \in
\mathcal{X}} p_{\mathbf{Y} | \mathbf{X}}(\mathbf{x} |
\mathbf{u})}\right)
\end{equation}
\noindent or, in a more explicit form:
\begin{equation*} \label{eq:hb0y4}
\begin{split}
H(B_{n,i}| \mathbf{Y} ) \leq - & \frac{2\sum_{\mathbf{x}
 \in \mathcal{X}_{n,i}(0)}\sum_{\mathbf{u} \in
\mathcal{X}_{n,i}(0)} p_{\mathbf{Y} | \mathbf{X}}(\mathbf{x} |
\mathbf{u})}{\sum_{\mathbf{x}
 \in \mathcal{X}}\sum_{\mathbf{u} \in
\mathcal{X}} p_{\mathbf{Y} | \mathbf{X}}(\mathbf{x} | \mathbf{u})} \times \\
& \log \frac{2\sum_{\mathbf{x}
 \in \mathcal{X}_{n,i}(0)}\sum_{\mathbf{u} \in
\mathcal{X}_{n,i}(0)} p_{\mathbf{Y} | \mathbf{X}}(\mathbf{x} |
\mathbf{u})}{\sum_{\mathbf{x}
 \in \mathcal{X}}\sum_{\mathbf{u} \in
\mathcal{X}} p_{\mathbf{Y} | \mathbf{X}}(\mathbf{x} | \mathbf{u})}
\\ - & \frac{2\sum_{\mathbf{x}
 \in \mathcal{X}_{n,i}(0)}\sum_{\mathbf{u} \in
\mathcal{X}_{n,i}(1)} p_{\mathbf{Y} | \mathbf{X}}(\mathbf{x} |
\mathbf{u})}{\sum_{\mathbf{x}
 \in \mathcal{X}}\sum_{\mathbf{u} \in
\mathcal{X}} p_{\mathbf{Y} | \mathbf{X}}(\mathbf{x} | \mathbf{u})} \times \\
& \log \frac{2\sum_{\mathbf{x}
 \in \mathcal{X}_{n,i}(0)}\sum_{\mathbf{u} \in
\mathcal{X}_{n,i}(1)} p_{\mathbf{Y} | \mathbf{X}}(\mathbf{x} |
\mathbf{u})}{\sum_{\mathbf{x}
 \in \mathcal{X}}\sum_{\mathbf{u} \in
\mathcal{X}} p_{\mathbf{Y} | \mathbf{X}}(\mathbf{x} | \mathbf{u})}
\end{split}
\end{equation*}

The above equation, substituted in \refeq{ih} and, then, back in
\refeq{cpcpm}, gives the lower bound on the pragmatic capacity
that is the starting point for our optimization.

\subsection{Mapping optimization}

The problem of mapping optimization has been reduced to the minimization of the
upper bound in \refeq{hb0y3} (or \refeq{hb0y4}). A binary entropy
function is minimized if its argument distribution is made as
unbalanced as possible. For large SNR, $ \sum_{\mathbf{x}
 \in \mathcal{X}_{n,i}(0)}\sum_{\mathbf{u} \in
\mathcal{X}_{n,i}(0)} p_{\mathbf{Y} | \mathbf{X}}(\mathbf{x},
\mathbf{u}) $ will be the dominant term in \refeq{hb0y3}, because it contains the
term with $\mathbf{x}=\mathbf{u}$; thus, it should be made as close
to 1 as possible. Conversely, the complementary term $
\sum_{\mathbf{x}
 \in \mathcal{X}_{n,i}(0)}\sum_{\mathbf{u} \in
\mathcal{X}_{n,i}(1)} p_{\mathbf{Y} | \mathbf{X}}(\mathbf{x},
\mathbf{u}), $ should be made as close to 0 as possible, i.e., it
should be minimized. We have come up with the following design
rule for the optimal mapping:
\begin{equation} \label{eq:mapr}
\mathcal{X}_{n,i}(0) = \arg \min_{\substack{\mathcal{X}' \subset
\mathcal{X} \\ |\mathcal{X}'| = |\mathcal{X}|/2}} \sum_{\mathbf{x}
 \in \mathcal{X}'}\sum_{\mathbf{u} \notin
\mathcal{X}'} p_{\mathbf{Y} | \mathbf{X}}(\mathbf{x}, \mathbf{u}).
\end{equation}

For practical reasons, the mapping is generated by an $m$-bit
\emph{labelling} of the trellis edges. Let $\mathcal{T}_{n,i}(0)$
($\mathcal{T}_{n,i}(1)$) be the subset of trellis edges at time
$n$ whose label has a 0 (1) in the $i$-th position. Then,
$\mathcal{X}_{n,i}(0)$ is constituted by all CPM waveforms whose
trellis paths pass through an edge belonging to
$\mathcal{T}_{n,i}(0)$, and the design rule in \refeq{mapr} can be
restated in terms of $\mathcal{T}_{n,i}(0)$.

We impose the condition that the trellis labelling must be
\emph{right-resolving}, i.e., edges leaving the same trellis state
have different binary labels. This condition does not force a
trellis labelling that varies with $n$. Moreover, the design
criterion in \refeq{mapr} is also independent of $n$. From these
facts, we deduce that \emph{the optimal mapping is generated by a
time-invariant trellis labelling}. Thus $\mathcal{T}_{n,i}(0) =
\mathcal{T}_{i}(0)$, for every $n$. Instead, the condition of
having a right-resolving trellis implies that
$\mathcal{T}_{1}(0),\dots,\mathcal{T}_{m}(0)$ cannot be chosen
independently, for $m > 1$.

In practice, we approximate the design rule in \refeq{mapr} with
the following suboptimal rule, which derives from the assumption
that, for large SNR, there is one dominant term in the sum of
\refeq{mapr}:
\begin{equation} \label{eq:mapr1}
\mathcal{X}_{n,i}(0) = \arg \min_{\mathcal{X}' \subset
\mathcal{X}} \left\{\max_{\mathbf{x}
 \in \mathcal{X}'}\max_{\mathbf{u} \notin
\mathcal{X}'} p_{\mathbf{Y} | \mathbf{X}}(\mathbf{x},
\mathbf{u})\right\}.
\end{equation}
Based on \refeq{mapr1}, our mapping optimization procedure
consists then of the following steps:
\begin{itemize}
\item For every pair of trellis edges, we apply the BCJR algorithm
to compute ${\max_{\mathbf{u}}} {\max_{\mathbf{x}}} p_{\mathbf{Y}
| \mathbf{X}}(\mathbf{x}, \mathbf{u})$ over all pairs of paths
passing through the given pair of edges at time zero. To do this
in general, we have to extend the trellis both at the left
(negative time instants) and at the right (positive time instants)
of the zeroth section. After a few trellis steps, the path metrics
reach a steady-state value, so there is no need to proceed
further.

\item The previous step yields a metric for each pair of edges.
Since we want to minimize the expression between braces in
\refeq{mapr1}, we partition the edges into $M = 2^m$ equal-size
clusters, corresponding to the $M$ different $m$-bit labels, in
such a way that pairs of edges with the highest metric are all
clustered together. In doing this, we force pairs of edges with
the same starting state into different clusters, to allow for a
right-resolving labelling.

\item We map the $M = 2^m$ clusters to $m$-tuples of bits
according to a Gray mapping. More precisely, we define the
distance between two clusters as the minimum distance between CPM
waveforms passing at time zero through edges belonging to the two
clusters. Cluster pairs with the smallest distance will be
associated to binary labels with Hamming distance 1.
\end{itemize}

The highest metric corresponds to edge pairs that belong to the
same trellis section of a minimum-distance error event. It may be
questionable whether it is possible to cluster \emph{all} pairs of
edges with that property, apart from those with the same starting
state, and whether this gives a \emph{unique} clustering. In the
next subsection, we will completely answer these questions for
$M=2$, and give the analytical expression of the optimal mapping,
provided that some necessary conditions are met.

\subsection{The optimal mapping for $M=2$}
\label{sec:pragcap}

As described in the previous subsection, the algorithm for mapping
optimization puts into the same cluster all edge pairs that belong
to the same trellis section of a minimum-distance error event. For
labelling purposes, we would like to obtain in this way exactly
$M$ clusters, but this happens only at certain conditions. In this
subsection, we derive such conditions for \emph{binary} CPM
schemes.

Let us consider a CPM scheme with $M=2$, impulse length $L$ and
modulation index $h=Q/P$. Trellis edges will be denoted hereafter
through the couple $(\mathbf{\alpha}, \beta)$, where
$\mathbf{\alpha}$ is a length-$L$ binary vector, including the
correlative state and the input symbol, and $\beta \in \{0,\dots,
P-1\}$ represents the phase state. Let $x_1(t)$ and $x_2(t)$ be
two CPM waveforms, corresponding to input symbol sequences
$\mathbf{a}_1$ and $\mathbf{a}_2$, respectively. It is well known
that the Euclidean distance between these two CPM waveforms only
depends on the \emph{difference} sequence $\mathbf{b} \triangleq
\mathbf{a}_1-\mathbf{a}_2$, with elements belonging to
$\{-1,0,1\}$.

Now, consider a given difference sequence $\mathbf{b}$ and define
a graph $\mathcal{G}(\mathbf{b}) = (\mathcal{V},
\mathcal{E}(\mathbf{b}))$, where:
\begin{itemize}
\item the vertex set $\mathcal{V}$ is in one-to-one correspondence
to the set of trellis edges for the considered CPM scheme.

\item the edge set $\mathcal{E}(\mathbf{b})$ is constructed in the
following way: two vertices are connected by an edge if and only
if the corresponding trellis edges have different starting states
and belong to the same trellis section of an error event generated
by $\mathbf{b}$.
\end{itemize}

Let $\mathcal{C}_0, \dots,
\mathcal{C}_{\mathcal{N}(\mathbf{b})-1}$ be the
$\mathcal{N}(\mathbf{b})$ connected components of the graph
$\mathcal{G}(\mathbf{b})$. The following theorem gives the
important properties of $\mathcal{C}_0, \dots,
\mathcal{C}_{\mathcal{N}(\mathbf{b})-1}$.

\begin{theorem} \label{th:1}
Consider a CPM scheme with $M=2$, impulse length $L$ and
modulation index $h=Q/P$. Given a difference sequence
$\mathbf{b}$, with length $\Delta(\mathbf{b})$, the following
facts about the graph $\mathcal{G}(\mathbf{b})$ defined above
hold:
\begin{enumerate}
\item If $\Delta(\mathbf{b}) = 2$, then $\mathcal{N}(\mathbf{b}) =
P$, otherwise $\mathcal{N}(\mathbf{b}) = 1$.

\item If $\mathcal{N}(\mathbf{b}) = P$, the connected component
$\mathcal{C}_i$, $i=0,\dots,P-1$ is the subgraph induced by the
subset of vertices\footnote{The notation $l_P$ means ``$l$ modulo
$P$''.}:
\begin{equation}
\mathcal{V}_i = \left\{\left(\mathbf{\alpha}, (i- Q
w_H(\mathbf{\alpha}))_P \right) : \mathbf{\alpha} \in \{0,1\}^L
\right\},
\end{equation}
independently of $\mathbf{b}$. All $\mathcal{C}_i$'s have size
$2^L$.

\end{enumerate}
\end{theorem}

\proof See Appendix \ref{sec:App1}.
\endproof

Suppose now that $\mathbf{b}$ gives the minimum Euclidean distance
of the CPM scheme. We want to partition the edges into two
clusters, namely, $\mathcal{T}(0)$ and $\mathcal{T}(1)$ (edges
with a label 0 and edges with a label 1, respectively). To meet
all constraints given by the graph $\mathcal{G}(\mathbf{b})$,
every subset $\mathcal{V}_i$ should be entirely contained into one
of the two clusters. This is clearly not possible if
$\Delta(\mathbf{b}) \neq 2$, because in that case
$\mathcal{V}_1=\mathcal{V}$.

If $\Delta(\mathbf{b}) = 2$, instead, we can place exactly $P/2$
subsets into $\mathcal{T}(0)$ and the other $P/2$ into
$\mathcal{T}(1)$. (Notice that the sizes of $\mathcal{T}(0)$ and
of $\mathcal{T}(1)$ both must be equal to $P2^{L-1}$.) This can be
done only if $P$ is even, otherwise, one of the $\mathcal{V}_i$'s
must be split in two. Thus, we have proved the following
proposition:
\begin{proposition} \label{prop:2}
Consider a CPM scheme with $M=2$, impulse length $L$ and
modulation index $h=Q/P$. We can fully perform the algorithm of
mapping optimization described in the previous section if and only
if:
\begin{itemize}
\item Every difference sequence $\mathbf{b}$ giving the minimum
Euclidean distance of the scheme has length two, and

\item $P$ is even.
\end{itemize}
\end{proposition}

If the conditions of Prop. \ref{prop:2} are satisfied, then, the
optimal mapping is uniquely determined by the condition of
right-resolving labelling. In fact, the two trellis edges leaving
a given trellis state will belong to $\mathcal{V}_i$ and
$\mathcal{V}_{(i+Q)_P}$, for some $i$. Thus, for the labelling to
be right-resolving, it must be:
\begin{equation}
\mathcal{T}(0) = \mathcal{V}_0 \cup \mathcal{V}_{(2Q)_P} \cup
\dots \cup \mathcal{V}_{((P-2)Q)_P}
\end{equation}
and
\begin{equation}
\mathcal{T}(1) = \mathcal{V}_{Q_P} \cup \mathcal{V}_{(3Q)_P} \cup
\dots \cup \mathcal{V}_{((P-1)Q)_P},
\end{equation}
or vice versa. In this way, all CPM waveform pairs at a minimum
distance between each other will be associated to binary input
sequence pairs with Hamming distance equal to 1.

If the conditions of Prop. \ref{prop:2} are not satisfied, then
the mapping optimization algorithm of the previous section cannot
be fully performed. It means that some CPM waveform pairs with
minimum distance will be associated to binary input sequence pairs
with Hamming distance larger than 1. A reasonable approach in this
case is to try to minimize the number of such CPM waveform pairs.

\subsection{The optimal mapping for $M>2$}

In non-binary CPM schemes, things become more involved. In general,
we have not found necessary and sufficient conditions to fully
perform the algorithm of mapping optimization.

Here, we give a generalization of Theorem \ref{th:1}, which
however deals only with a subset of possible difference sequences.
\begin{theorem} \label{th:2}
Consider a CPM scheme with $M>2$, impulse length $L$ and
modulation index $h=Q/P$. Let
$\mathbf{b}=(b_1,\dots,b_{\Delta(\mathbf{b})})$  be a difference
sequence  with length $\Delta(\mathbf{b})$ and elements in
$\{0,\pm 1, \dots, \pm (M-1)\}$. Let $b_{\Delta(\mathbf{b})} = \pm
1$. The following facts about the graph $\mathcal{G}(\mathbf{b})$
defined above hold:
\begin{enumerate}
\item If $\Delta(\mathbf{b}) = 2$, then $\mathcal{N}(\mathbf{b}) =
P$, otherwise $\mathcal{N}(\mathbf{b}) = 1$.

\item If $\mathcal{N}(\mathbf{b}) = P$, the connected component
$\mathcal{C}_p$, $p=0,\dots,P-1$ is the subgraph induced by the
subset of vertices
\begin{equation} \label{eq:Vi}
\begin{split}
\mathcal{V}_p = & \left\{\left(\mathbf{\alpha}, \left(p- Q
\sum_{l=1}^L a_l\right)_P \right) : \right. \\
& \left. \mathbf{\alpha}
=(a_1,\dots,a_L) \in \{0,\dots,M-1\}^L \right\},
\end{split}
\end{equation}
independently of $\mathbf{b}$. All $\mathcal{C}_p$'s have size
$M^L$.

\end{enumerate}
\end{theorem}

\proof See Appendix \ref{sec:App2}.
\endproof

Following the same arguments of the previous subsection, we thus
obtain the following proposition, which gives only sufficient (not
necessary) conditions to fully perform the mapping optimization
algorithm.

\begin{proposition} \label{prop:4}
Consider a CPM scheme with $M>2$, impulse length $L$ and
modulation index $h=Q/P$. We can fully perform the algorithm of
mapping optimization described in the previous section if:
\begin{itemize}
\item Every difference sequence $\mathbf{b}$ giving the minimum
Euclidean distance of the scheme has length two and $b_2 = \pm 1$,
and

\item $P$ is a multiple of $M$.
\end{itemize}
\end{proposition}

If the conditions of Prop. \ref{prop:4} are satisfied, then, by
constraining the  resulting labelling to be right-resolving, we
obtain the following $M$ clusters:
\begin{equation} \label{eq:Ti}
\widetilde{\mathcal{T}}(i) = \mathcal{V}_{(iQ)_P} \cup
\mathcal{V}_{((M+i)Q)_P} \cup \dots \cup
\mathcal{V}_{((P-M+i)Q)_P},
\end{equation}
for $i=0,\dots,M-1$.

Finally, the clusters $\widetilde{\mathcal{T}}(i)$'s are mapped to
the $M$ binary labels according to Gray mapping.

In the cases that are not covered by the hypotheses of Theorem
\ref{th:2}, the algorithm may or may not be fully performed
depending on the scheme parameters and on the difference sequences
$\mathbf{b}$ giving the minimum Euclidean distance. If the
algorithm can be fully performed, in general, the $M$ clusters may
not be as in \refeq{Ti}.

%%%%%%%%%%%%%%%%%%%%%%%%%%%%%%%%%%%%%%%%%%%%%%%%%%%%%%%%%%%%%%%%%%%%%%
\subsection{The optimized CPE}
%%%%%%%%%%%%%%%%%%%%%%%%%%%%%%%%%%%%%%%%%%%%%%%%%%%%%%%%%%%%%%%%%%%%%%

Starting from (\ref{eq:Vi}), where we define

\[
a_{L+1} = \left(p - Q \sum_{l=1}^L{a_l}\right)_P
\]

\noindent we derive the input symbol $a_1$ to the CPE as the solution
to the following equation

\[
(Q a_1)_P = \left(p - a_{L+1} - Q \sum_{l=2}^L{a_l}\right)_P.
\]

\noindent Since $P$ and $Q$ are relatively prime, such solution is unique.
Moreover, from (\ref{eq:Vi}) we have $a_1 \in \{0, \ldots, M - 1\}$ and,
since also $M$ and $Q$ are relatively prime,  $(Q a_1)_M$ entails a unique
solution for $a_1$.
As a result, we can write

\begin{equation}\label{eq:Qa1}
(Q a_1)_M = \left(p - a_{L+1} - Q \sum_{l=2}^L{a_l}\right)_M.
\end{equation}

Clustering as in (\ref{eq:Ti}) can be performed observing that the vertex subset indices
in each cluster can be written as $p = ((kM + i) Q)_P$.   Inserting this in (\ref{eq:Qa1})
and recalling that $P$ is a multiple of $M$ we obtain

\begin{equation*}\label{eq:Qa1-2}
\begin{split}
(Q a_1)_M & = \left(iQ + k M Q - a_{L+1} - Q \sum_{l=2}^L{a_l}\right)_M \\
& = \left(iQ - a_{L+1} - Q \sum_{l=2}^L{a_l}\right)_M.
\end{split}
\end{equation*}

\noindent where $i$ is the cluster label. Applying the distributive property of multiplication
over addition in the ring $\mathbb{Z}_M$  we obtain

\begin{equation}
a_1 = \left(i - (a_{L+1}Q^{-1})_M - \sum_{l=2}^L{a_l}\right)_M.
\label{eq:a1f}
\end{equation}

\begin{figure}[!ht]
\centering
\includegraphics[width=3.5in]{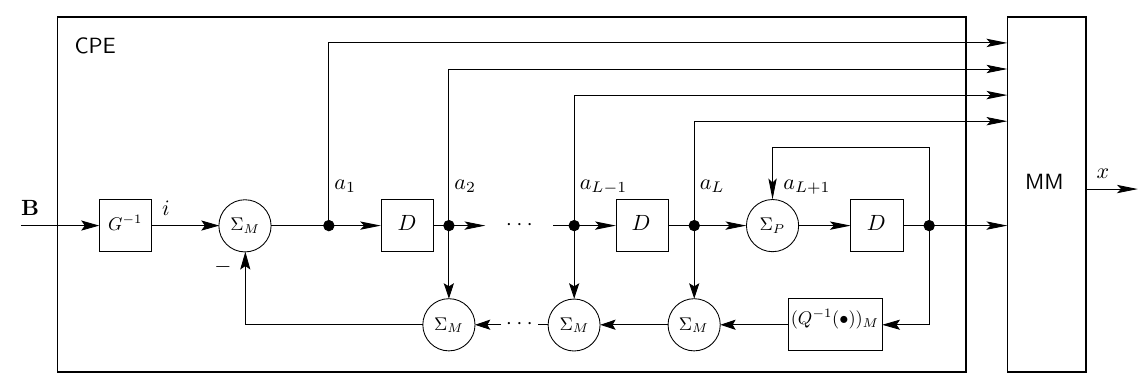}
\caption{The optimized CPM encoder. The block labelled
$\Sigma_P$ (resp. $\Sigma_M$) is a modulo-$P$ (resp. modulo-$M$) adder.
The block labelled $G^{-1}$ is the inverse of the Gray function defined in~\cite[ch.20]{bib:nr}.}
\label{fig:cpmencopt}
\end{figure}

\noindent Finally, the binary label $\bf B$ is obtained as
$\mathbf{B} = G(i)$, where $G$ is the Gray function defined in~\cite[ch.20]{bib:nr}.
Fig.~\ref{fig:cpmencopt} shows the scheme
of the optimized CPM encoder.

Adopting the optimized CPE, i.e., starting from the CPE structure of Fig.~\ref{fig:cpmencopt},
a procedure similar to the one described in Sec.~\ref{sec:opt} has been
performed in order to select the CPM schemes with best $C_{\rm P-CPM}$.
Results are shown in Tab.~\ref{tab:bestrec99p} and Tab.~\ref{tab:bestrc99p}.

\begin{table*}[ht]
\centering\includegraphics[angle=-90,width=7in]{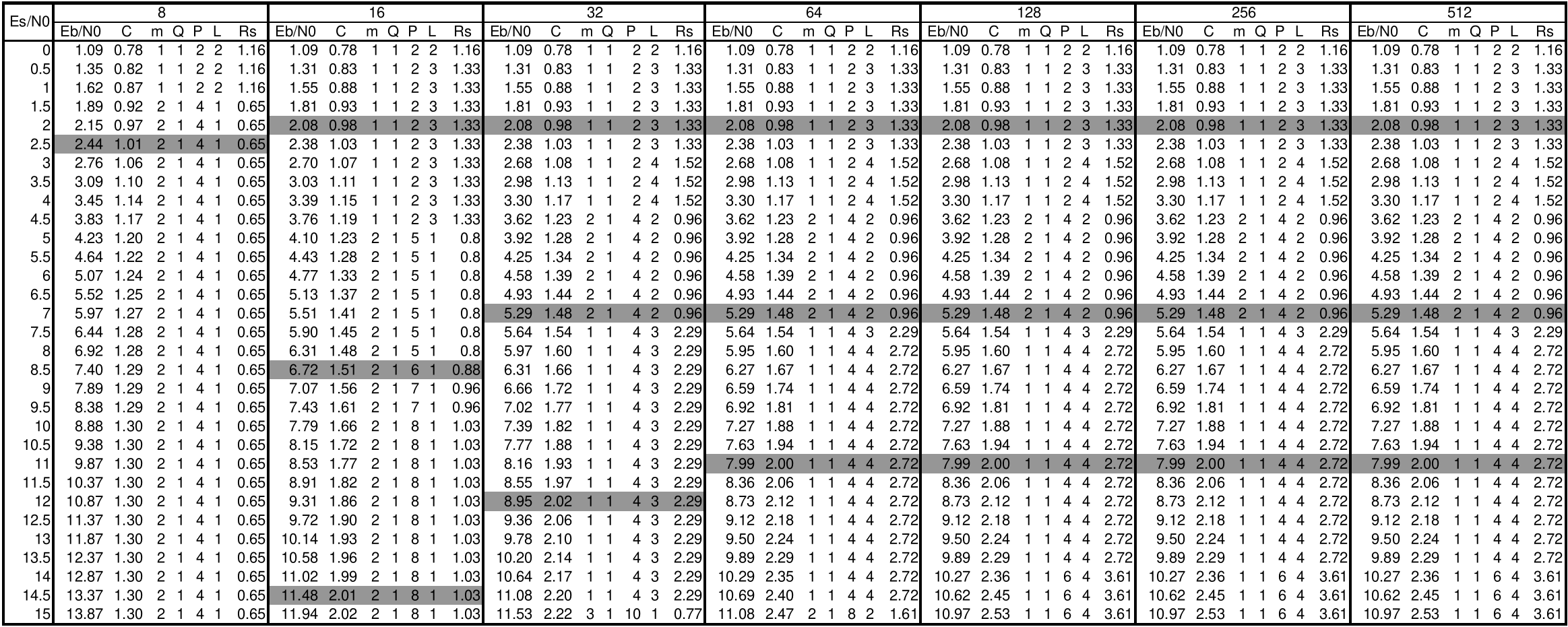}
\caption{Table of the P-CPM schemes with best $C_{\rm P-CPM}$ and
rectangular frequency pulse at 99\% bandwidth. The
complexity range is 8 to 512. The highlighted entries
correspond to the selected CPM schemes for target capacities of 1, 1.5 and 2 bits/s/Hz.}
\label{tab:bestrec99p}
\end{table*}

\begin{table*}[ht]
\centering\includegraphics[angle=-90,width=7in]{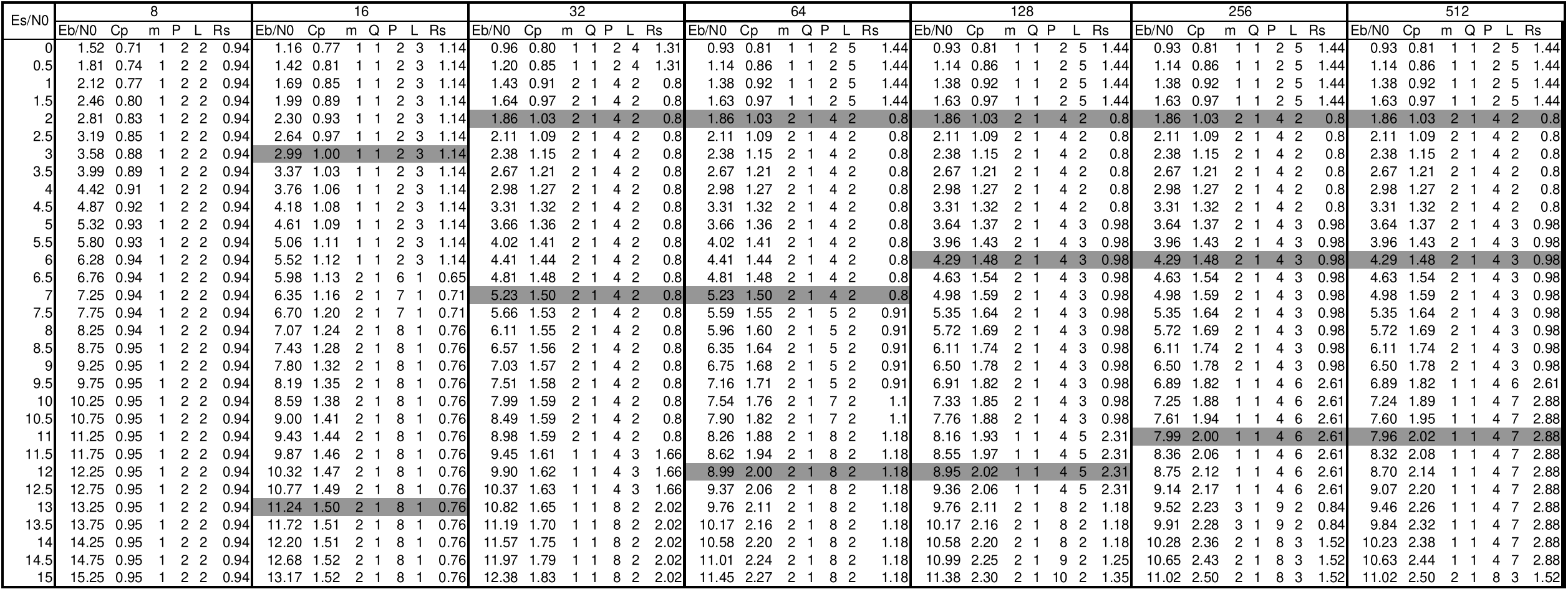}
\caption{Table of the P-CPM schemes with best $C_{\rm P-CPM}$ and
raised-cosine frequency pulse at 99\% bandwidth. The complexity range
is 8 to 512. The highlighted entries
correspond to the selected CPM schemes for target capacities of 1, 1.5 and 2 bits/s/Hz.}
\label{tab:bestrc99p}
\end{table*}

A comparison of Tab.~\ref{tab:bestrec99p} with Tab.~\ref{tab:bestrec99} and
Tab.~\ref{tab:bestrc99p} with Tab.~\ref{tab:bestrc99} highlights that, in some
cases, the CPM schemes with best $C_{\rm CPM}$ feature the best  $C_{\rm P-CPM}$,
as is the case, e.g., for REC pulses with complexity 64 and capacity 1.5 bits/s/Hz
and for RC pulses with complexity 64 and capacity 1.5 bits/s/Hz and 2 bits/s/Hz.
In some other cases, the selected schemes do not match.

We observed that, in many binary CPM schemes, the pragmatic capacity resulting from the
CPE optimization approaches the CPM capacity closer than for non binary schemes.
As a result, even when featuring a worse CPM capacity, these schemes may exhibit a larger
pragmatic capacity.
This is the case, e.g., of REC pulses with complexity 64 and capacity 1 bits/s/Hz, where
the scheme with best $C_{\rm CPM}$ is quaternary and the scheme with best $C_{\rm P-CPM}$
is binary.

The optimal mapping described above results in a CPM scheme with finite $d_{E1}$,
the minimum Euclidean distance at the CPM output corresponding to an input Hamming
distance 1. To prove this, we use the polynomial notation with dummy variable $D$ for 
symbol sequences and we observe from Fig.~\ref{fig:cpmencopt} that

\begin{equation}
a_{i+1}(D) = D a_i(D), \quad i = 1, \ldots , L - 1
\label{eq:aip1}
\end{equation}

\noindent and

\begin{equation}
a_{L+1}(D) = \left(a_L(D) \frac{D}{1 - D}\right)_P
\label{eq:alp1}
\end{equation}

\noindent If $Q = 1$, from (\ref{eq:alp1}), (\ref{eq:aip1}) and (\ref{eq:a1f}), we obtain

\[a_1(D) = \left(i(D) (1 - D)\right)_M.\]

\noindent Therefore $a_1(D)$ can be obtained by feedforward encoding
of $i(D)$.   From (\ref{eq:aip1}) it is straightforward to
deduct that all $a_i(D)$, $i = 1, \ldots, L$ have the same property.

As for symbol $a_{L+1}(D)$, from (\ref{eq:alp1}) we obtain

\begin{equation}
a_{L+1}(D) = \left(\frac{D^L a_1(D)}{1 - D}\right)_P = \left(\frac{D^L \left(i(D) (1 - D)\right)_M}{1 - D}\right)_P
\label{eq:alp1-1}
\end{equation}

\noindent If $M = z P, z \in \mathbb{N}$, this clearly proves that symbol $a_{L+1}(D)$ can be 
obtained by feedforward encoding $i(D)$.
When $P = w M, w \in \mathbb{N}$, it is possible to rewrite (\ref{eq:alp1-1}) as

\begin{eqnarray}
\nonumber a_{L+1}(D) & = & \left(\frac{D^L \left(i(D) (1 - D)\right)_M}{1 - D}\right)_M \\ 
\nonumber & + & M \left\lfloor\frac{D^L \left(i(D) (1 - D)\right)_M}{M (1 - D)}\right\rfloor_{P/M} \\
\nonumber & = & \left(D^L i(D)\right)_M \\
& + & M \left\lfloor\frac{D^L \left(i(D) (1 - D)\right)_M}{M (1 - D)}\right\rfloor_{P/M}
\label{eq:alp1-2}
\end{eqnarray}

\noindent When the last term of (\ref{eq:alp1-2}) is zero, $a_{L+1}(D)$ can be obtained by
feedforward encoding from $i(D)$, therefore the encoder is feedforward.
For \emph{difference} sequences $i(D)$ with length 1 and symbols in $\{-1, +1\}$ this condition is met.
Moreover, due to Gray mapping, the corresponding pairs of binary input sequences have unitary
Hamming distance, hence the resulting encoder has finite $d_{E1}$.

It is well known that in serially concatenated schemes, in order to take advantage
of the interleaver gain~\cite{bib:Benedetto-Unveiling}, the inner constituent encoder
or modulator must have infinite $d_{E1}$~\cite{bib:ett}.
For this reason, when used in SC-CPM schemes, the optimal mapping leads to poor
performance and hence a recursive CPE must be employed.
We have thus used the mapping induced by the Rimoldi decomposition, which 
has this property.

%%%%%%%%%%%%%%%%%%%%%%%%%%%%%%%%%%%%%%%%%%%%%%%%%%%%%%%%%%%%%%%%%%%%%%
\section{Implementation of the optimized coded CPM schemes}
\label{sec:implementation}
%%%%%%%%%%%%%%%%%%%%%%%%%%%%%%%%%%%%%%%%%%%%%%%%%%%%%%%%%%%%%%%%%%%%%%

Using the procedures described in the previous sections, SC-CPM and P-CPM schemes have
been designed with target capacities of $C_T = 1$ bits/s/Hz, $C_T = 1.5$ bits/s/Hz and
$C_T = 2$ bits/s/Hz.

%%%%%%%%%%%%%%%%%%%%%%%%%%%%%%%%%%%%%%%%%%%%%%%%%%%%%%
\subsection{Selection of the modulation schemes}

\begin{figure}[ht]
\centering
\includegraphics[angle=270,width=3.5in]{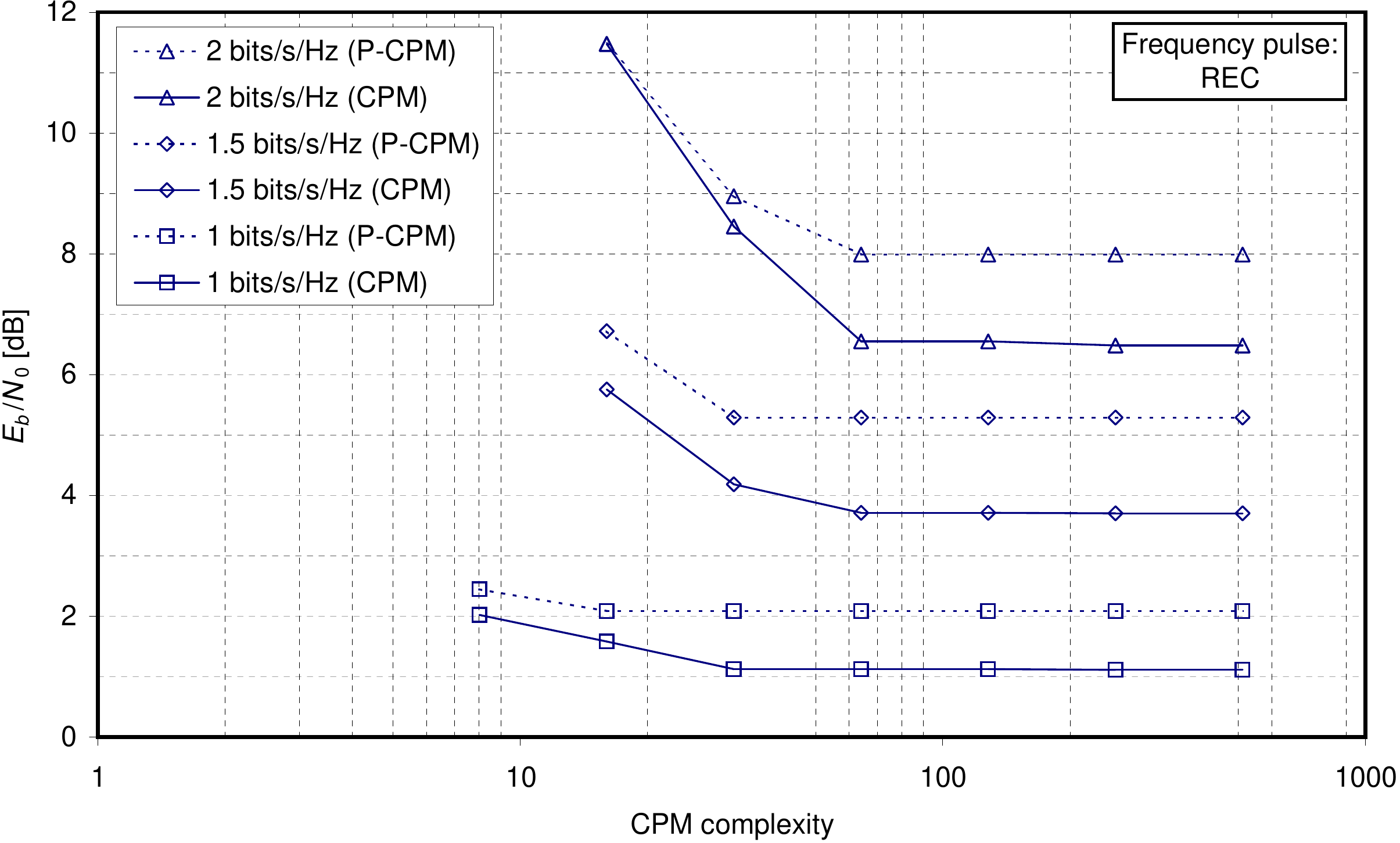}
\caption{Energy efficiency of REC CPM schemes.}
\label{fig:PowerEffREC}
\end{figure}

\begin{figure}[ht]
\centering
\includegraphics[angle=270,width=3.5in]{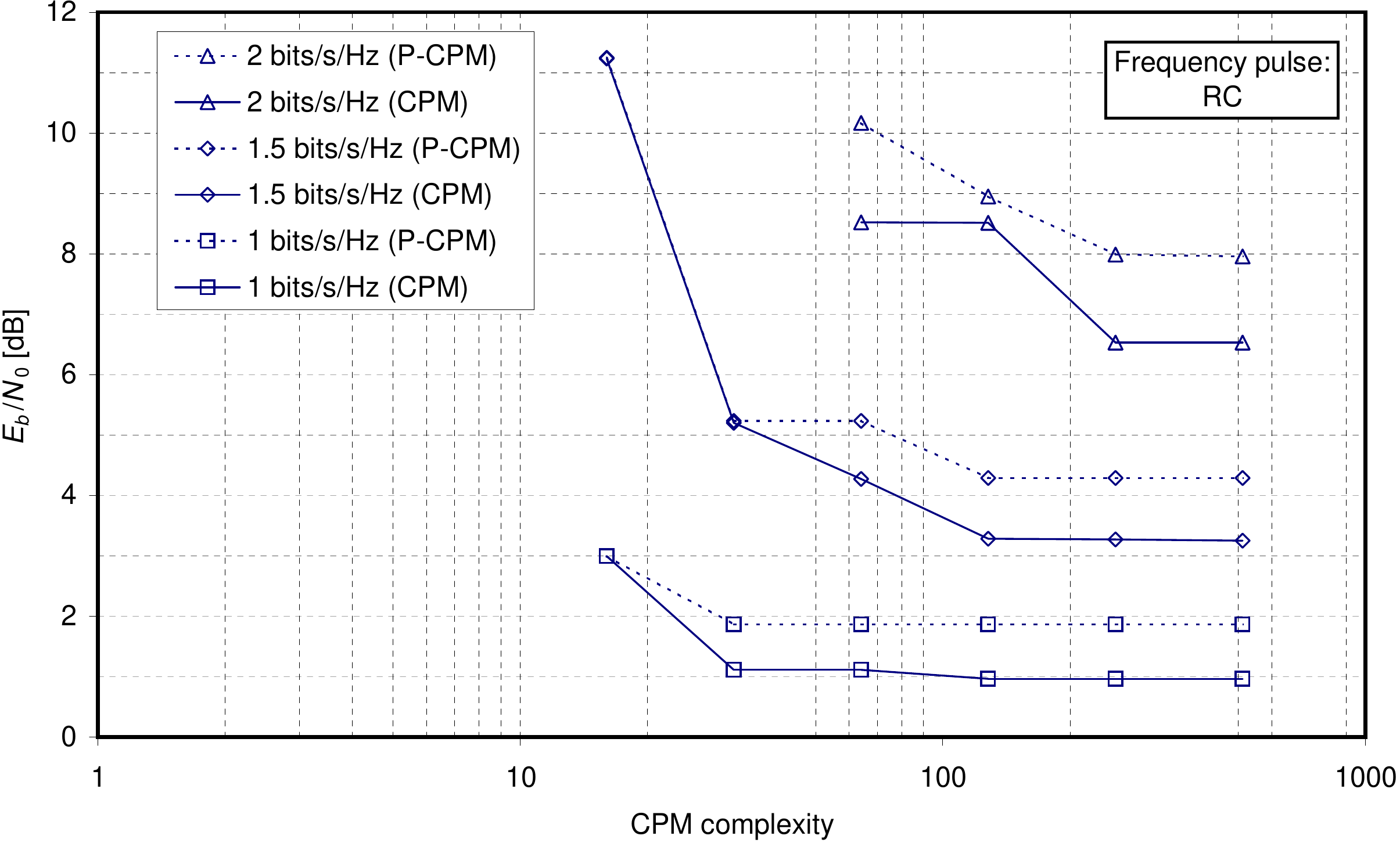}
\caption{Energy efficiency of RC CPM schemes.}
\label{fig:PowerEffRC}
\end{figure}

Fig.~\ref{fig:PowerEffREC} and Fig.~\ref{fig:PowerEffRC} show the $E_b/N_0$
values plotted against the CPM complexity for different target capacities.
Continuous curves refer to CPM capacity and dashed curves refer to pragmatic
capacity.

For REC schemes (Fig.~\ref{fig:PowerEffREC}) we observe that increasing the
complexity beyond 64 yields a negligible gain both in CPM capacity and in pragmatic
capacity. Hence, CPM schemes with REC frequency pulse and maximum
complexity 64 have been considered.

For RC schemes (Fig.~\ref{fig:PowerEffRC}), increasing the
complexity beyond 64 yields a significant gain both in CPM capacity and in pragmatic
capacity at 1.5 bits/s/Hz and 2 bits/s/Hz, hence CPM schemes with higher complexities
should be chosen. However, in SC-CPM receivers, the increased complexity would be
enhanced by the iterative process, resulting in an impractical solution.

%The maximum CPM complexity has been set to 64 edges per CPM input bit.
%This value has been chosen considering that, as shown in Fig.~\ref{fig:PowerEffREC}
%for REC schemes, increasing the complexity beyond 64 yields a negligible gain.
%Even for RC schemes, which feature a significant gain even for CPM complexities beyond 128
%(see Fig.~\ref{fig:PowerEffRC}), the same complexity value has been adopted, since
%further increasing the complexity would significantly enhance the overall complexity and thus
%result in longer simulations.

%%%%%%%%%%%%%%%%%%%%%%%%%%%%%%%%%%%%%%%%%%%%%%
\subsection{Serially-concatenated schemes}

A serially concatenated CPM scheme consists of the cascade of an outer convolutional
encoder connected to the CPM modulator through an interleaver.

\begin{figure}
\centering
\includegraphics[clip,width=3in]{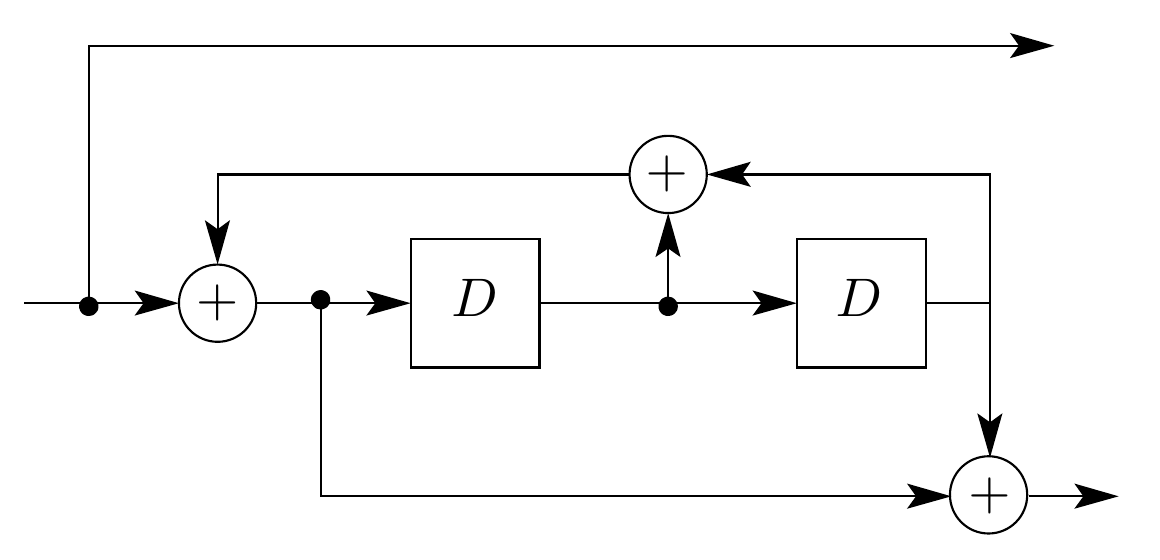}
\caption{Convolutional encoder used in SC-CPM schemes.}
\label{fig:convenc}
\end{figure}

As outer encoder, a 4-state, rate 1/2 systematic recursive convolutional encoder has been chosen. It is shown in Fig.~\ref{fig:convenc}. The choice has been dictated by a trade-off between performance and complexity. The encoder output can be punctured to achieve rates up to 3/4 while keeping a free distance larger than 2.

In order to obtain rates larger than 1/2, the output of the convolutional
encoder is punctured according to a rate-matching algorithm that selects
all systematic bits and some coded bits.  The puncturing rate is

\[
R_P = \frac{N_O}{N_I} \geq 1
\]

\noindent where $N_O$ and $N_I$ are two integers ($N_I \leq N_O \leq 2 N_I$).
The puncturing algorithm selects $N_I$ coded bits from each block of $N_O$ with
the following constraint: all systematic bits are selected; some coded bits
are selected to achieve the desired rate; the other bits are punctured.
The resulting outer code rate is

\[
R_{CC} = \frac{1}{2}\frac{N_O}{N_I}
\]

The interleaver that connects the punctured outer encoder to the inner CPM modulator
is a spread interleaver.

The design of a SC-CPM scheme consists in the choice of the CPM scheme and of the
outer encoder based on the desired capacity and on design constraints (e.g., the
maximum affordable complexity of the CPM scheme).

\begin{example}
We describe the design of a SC-CPM scheme with target capacity
$C_T = 1.5$ bits/s/Hz, REC frequency pulse and maximum CPM complexity of 64 edges per CPM
input bit.
As for SC-CPM schemes, the parameters that maximize $C_{\rm CPM}$ must be chosen,
from Tab.~\ref{tab:bestrec99}.
Moreover, since the maximum CPM complexity is 64, we must restrict the search to the
column set labelled 64.
Next, in the column labelled $C$, we look for the value closest to the target capacity.
We find that the capacity value closest to $C_T$ is 1.51 and the
corresponding CPM parameters are $m = 2$, $h = 1/5$, $L = 2$.
Such scheme exhibits a symbol rate $R_s = 1.18$ and a corresponding bit rate $R_b = m R_s = 2.36$.
The scheme of the CPM modulator corresponding to these parameters is shown in Fig.~\ref{fig:cpmenc-2-5-2}.

\begin{figure}[ht]
\centering
\includegraphics[width=3.5in]{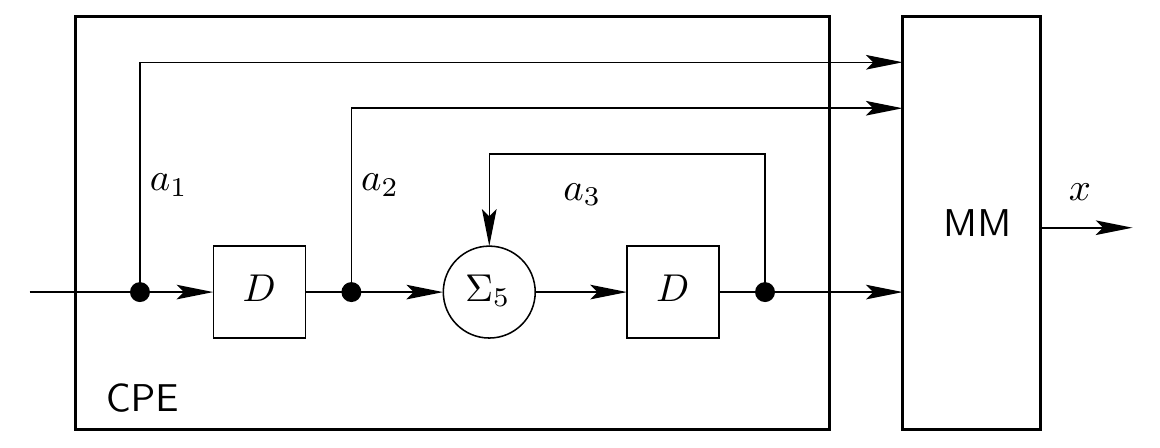}
\caption{Scheme of the CPM modulator with $m = 2$, $h = 1/5$, $L = 2$.}
\label{fig:cpmenc-2-5-2}
\end{figure}

In order to obtain a target capacity $C_T = 1.5$ bits/s/Hz, the outer code rate must be set to

\[
R_{CC} = \frac{N_O}{2 N_I} = \frac{C_T}{R_b} = 0.636
\]

\noindent therefore

\[
\frac{N_O}{N_I} = \frac{2 C_T}{R_b} = 1.2712 \simeq \frac{75}{59}.
\]

\noindent The final step consists in choosing two integers for $N_O$ and $N_I$ in order
to approximate with sufficient precision this ratio.

\flushright \QEDopen
\end{example}

The resulting design parameters for all considered target capacities are shown in Tab.~\ref{tab:c-cpm-designs}.

%%%%%%%%%%%%%%%%%%%%%%%%%%%%%%%%%%%%%%%%%%
\subsection{Pragmatic schemes}

A pragmatic CPM scheme consists of the cascade of a SCCC encoder and a CPM modulator.

The chosen encoder is the SCCC encoder described in~\cite{bib:MHOMS}.
The constituent convolutional encoders used in such scheme are the 4-state convolutional
encoders shown in Fig.~\ref{fig:convenc}.
The outer convolutional encoder is punctured to a rate 2/3.

The inner encoder is punctured using a carefully designed rate-matching algorithm
that selects all the systematic bits and some coded bits in order to achieve the desired
overall rate.

The interleaver that connects the punctured outer encoder to the inner encoder is a
spread interleaver.

\begin{example}
We show how to design  a P-CPM scheme with target capacity
of 2 bits/s/Hz, RC frequency pulse and maximum CPM complexity of 64 edges per CPM
input bit.
For P-CPM schemes, the inner CPM modulation parameters that maximize $C_{\rm P-CPM}$
must be chosen. Therefore, Tab.~\ref{tab:bestrc99p} must be considered. Moreover, since
the maximum CPM complexity is 64, we must restrict the search to the column set labelled 64.
Next, in column labelled $C$ we look for the value closest to the target capacity.
We find that the capacity value closest to $C_T$ is 2.00 and the
corresponding CPM parameters are $m = 2$, $h = 1/8$, $L = 2$.
Such scheme exhibits a symbol rate $R_s = 1.18$ and a corresponding bit rate
$R_b = m R_s = 2.36$.
The scheme of the CPM modulator corresponding to these parameters is shown in Fig.~\ref{fig:cpmencoptq-2-8-2}.

\begin{figure}[ht]
\centering
\includegraphics[width=3.5in]{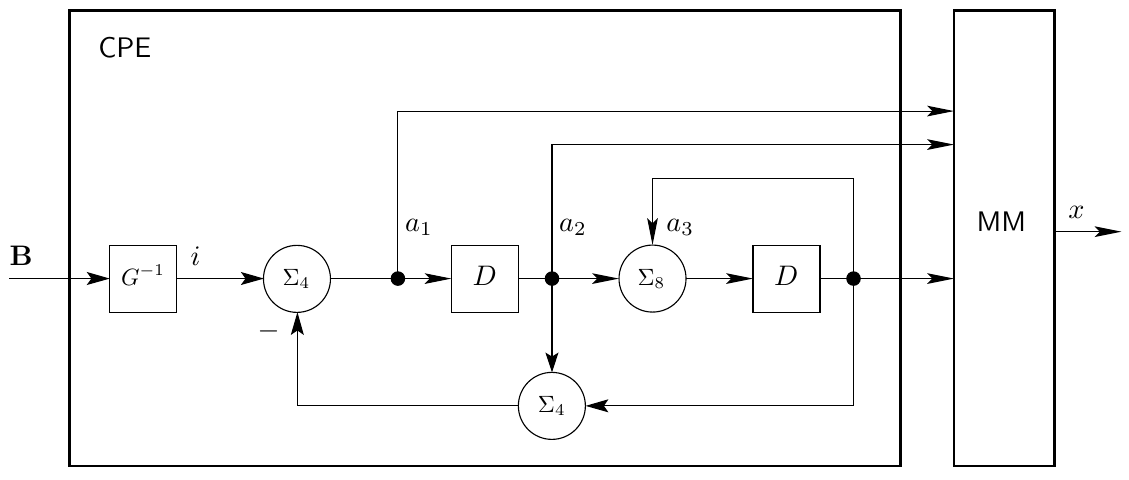}
\caption{Scheme of the CPM modulator with $m = 2$, $h = 1/8$, $L = 2$.}
\label{fig:cpmencoptq-2-8-2}
\end{figure}

In order to obtain a target capacity $C_T = 2$ bits/s/Hz, the outer code rate must be

\[
R_{\rm SCCC} = \frac{C_T}{R_b} = 0.847.
\]

\flushright \QEDopen
\end{example}

Tab.~\ref{tab:c-cpm-designs} summarizes the parameters of the designed schemes.

\begin{table}
\centering
\begin{tabular}{r|c||cc|ccc||c|ccc}
    & \multirow{2}{*}{$C_T$} & \multicolumn{5}{c||}{SC-CPM}  &  \multicolumn{4}{c}{P-CPM}   \\
                     &  & $N_O$ & $N_I$ & $m$ & $P$ & $L$ & $R_{\rm SCCC}$ & $m$ & $P$ & $L$      \\
\hline \hline
\multirow{3}{*}{\begin{sideways}REC\end{sideways}} &   1   &  25   &   24  &  2  &  4  &  2  &       0.752    &  1  &  2  &  3  \\
                     &   1.5 &  75   &   59  &  2  &  5  &  2  &       0.781    &  2  &  4  &  2  \\
                     &   2   &  40   &   27  &  2  &  6  &  2  &       0.735    &  1  &  4  &  4  \\
\hline
\multirow{3}{*}{\begin{sideways}RC\end{sideways}}  &   1   &    5  &    4  &  2  &  4  &  2  &       0.625    &  2  &  4  &  2  \\
                     &   1.5 &  150  &   91  &  2  &  5  &  2  &       0.938    &  2  &  4  &  2  \\
                     &   2   &  100  &   59  &  2  &  8  &  2  &       0.847    &  2  &  8  &  2  \\
\end{tabular}
\caption{Parameters of the optimized SC-CPM and P-CPM schemes yielding 1.0, 1.5 and 2.0 bits/s/Hz.}
\label{tab:c-cpm-designs}
\end{table}

%%%%%%%%%%%%%%%%%%%%%%%%%%%%%%%%%%%%%%%%%%%%%%%%%%%%%%%%%%%%%%%%%%%%%%
\section{Results}
\label{sec:results}
%%%%%%%%%%%%%%%%%%%%%%%%%%%%%%%%%%%%%%%%%%%%%%%%%%%%%%%%%%%%%%%%%%%%%%

\begin{figure}[ht]
\centering
\includegraphics[angle=270,width=3.5in]{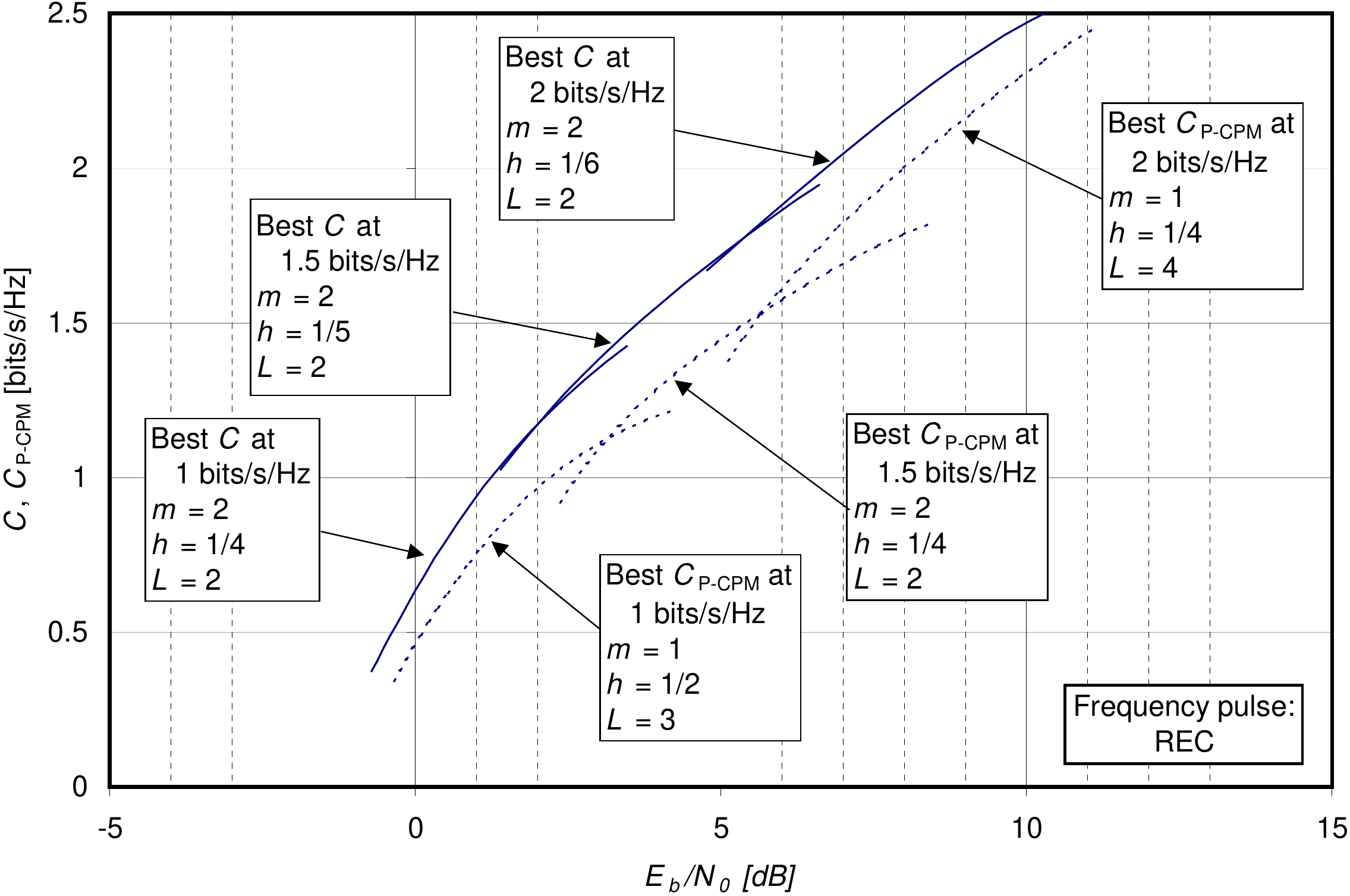}
\caption{CPM capacity and pragmatic capacity of
the selected CPM schemes with complexity $\mathcal{Y} \leq 64$. The
frequency pulse is REC.}
\label{fig:C64se-REC}
\end{figure}

\begin{figure}[ht]
\centering
\includegraphics[angle=270,width=3.5in]{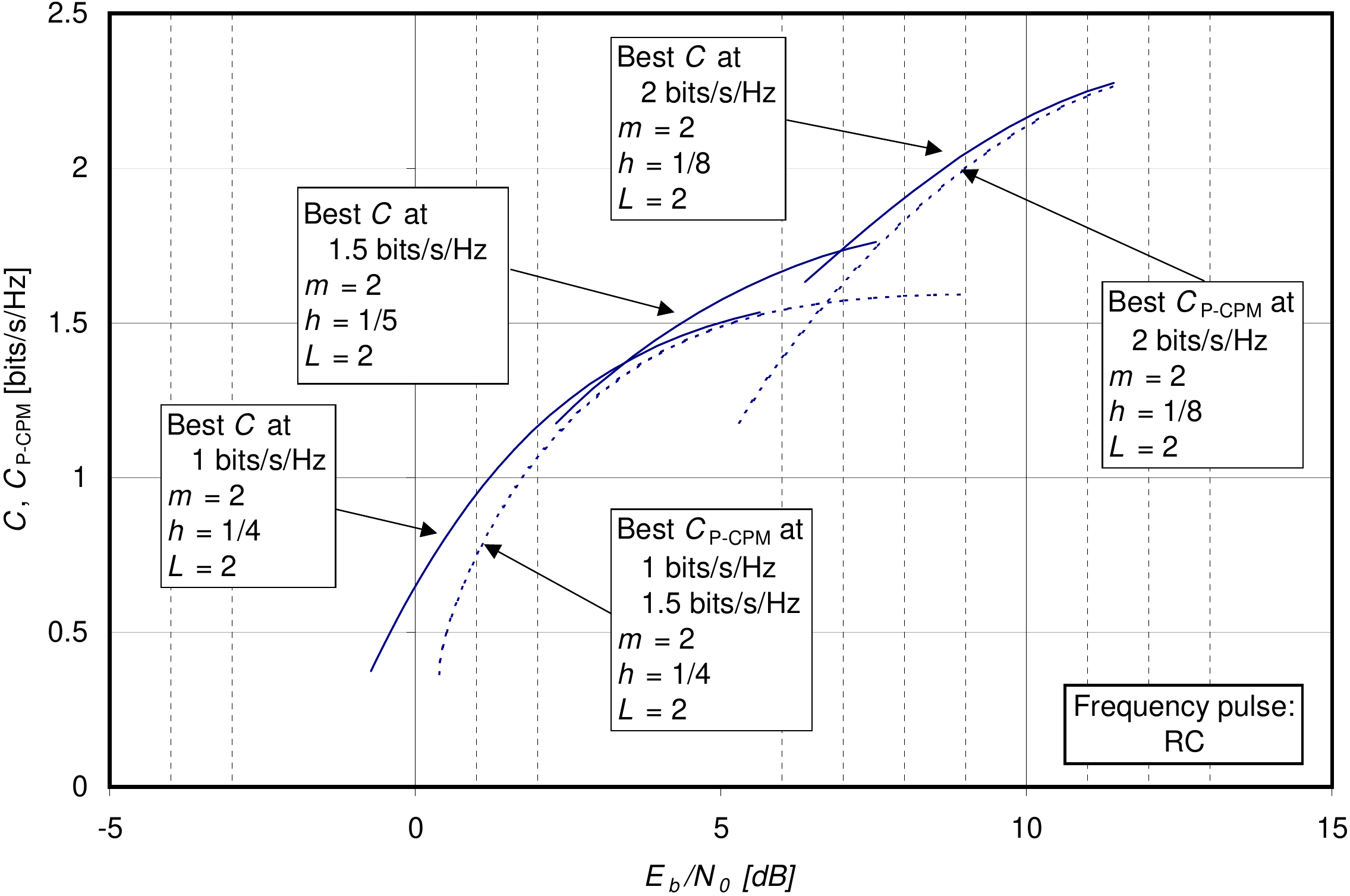}
\caption{CPM capacity and pragmatic capacity of
the selected CPM schemes with complexity $\mathcal{Y} \leq 64$. The
frequency pulse is RC.}
\label{fig:C64se-RC}
\end{figure}

The performance of the designed coding schemes reported in Tab.~\ref{tab:c-cpm-designs}
has been assessed through simulation.  For each scheme, the CPM capacity and the pragmatic
capacity have been estimated. The obtained results are shown in
Fig.~\ref{fig:C64se-REC} for REC frequency pulses and
Fig.~\ref{fig:C64se-RC} for RC frequency pulses.

%The region of achievable information rates for an SC-CPM scheme is bounded
%by the CPM capacity curve $C$ of its CPM scheme.
%\emph{CPM} spectral
%efficiency of the CPM scheme, therefore, the CPM schemes with largest
%\emph{CPM} spectral 0 have been chosen.

In both the P-CPM and the SC-CPM case, an outer information word
length of $K = 9840$ bits has been considered and the code rates have
been set according to Tab.~\ref{tab:c-cpm-designs}.
The number of decoding iterations has been set to 10.

\begin{figure}[!ht]
\centering
\includegraphics[angle=270,width=3.5in]{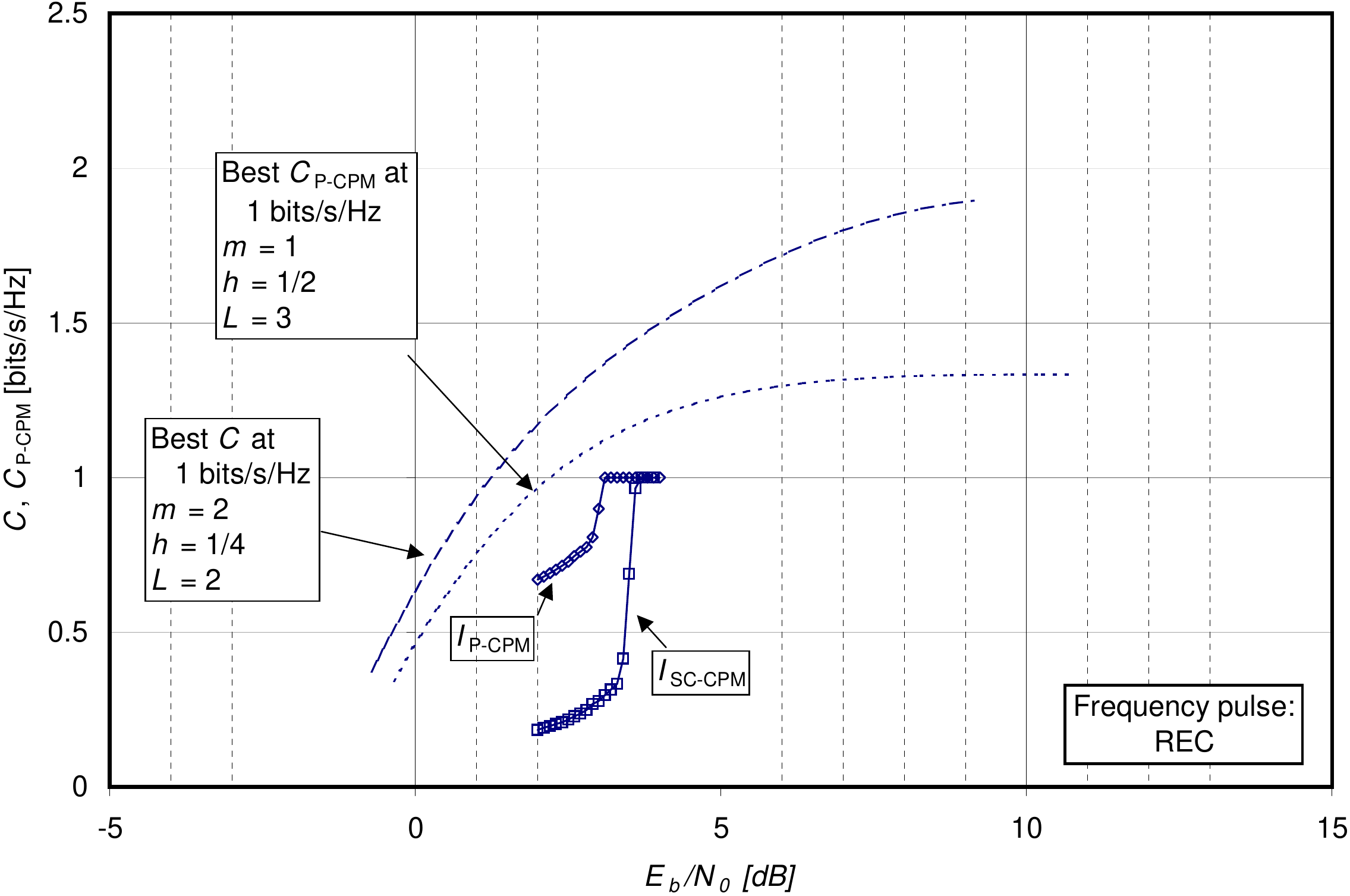}
\caption{Information rates of P-CPM and SC-CPM coded systems at 1
bits/s/Hz. ${\cal Y} = 64$, REC frequency pulse.}
\label{fig:C64se-1-REC}
\end{figure}

\begin{figure}[!ht]
\centering
\includegraphics[angle=270,width=3.5in]{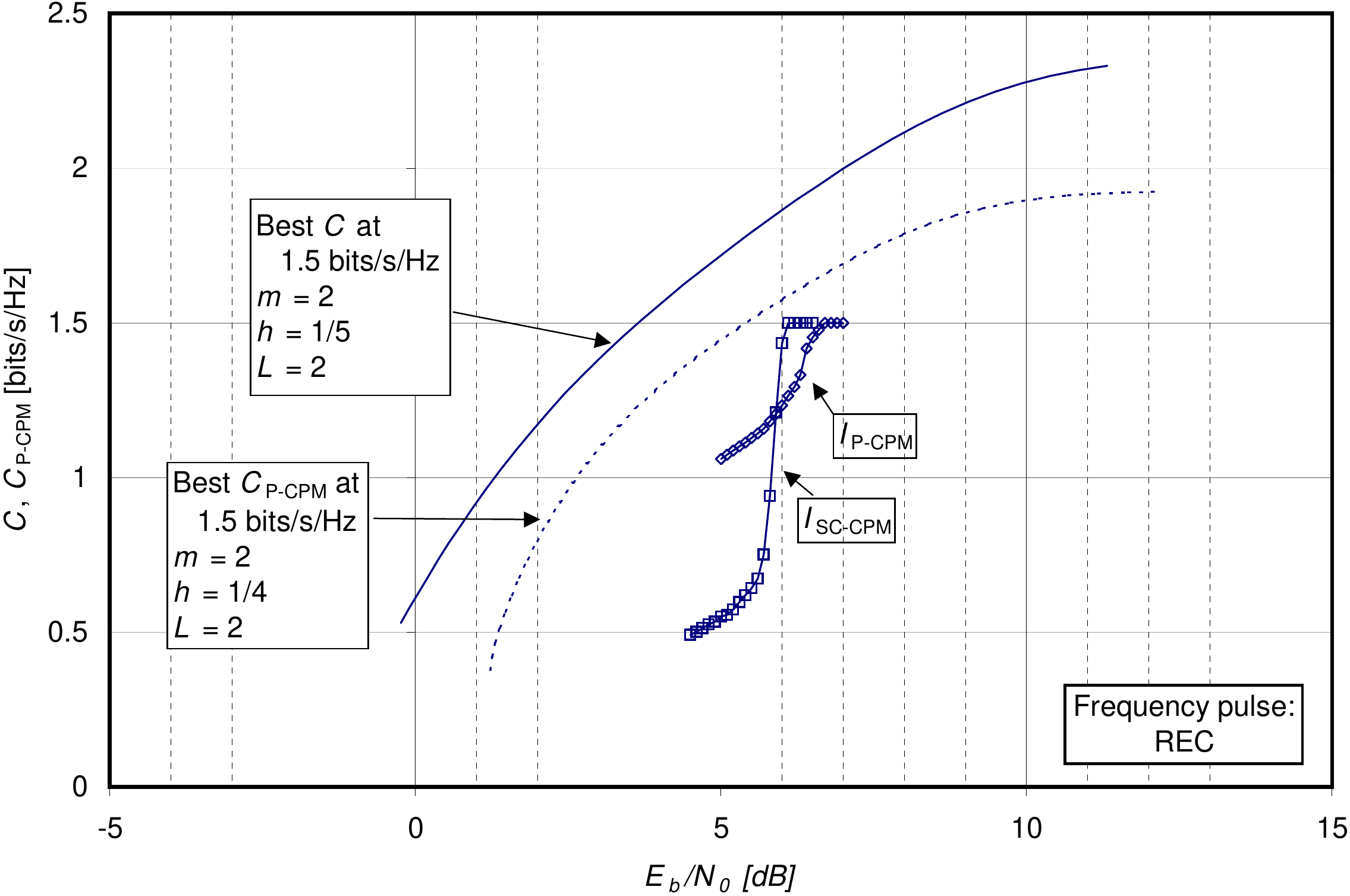}
\caption{Information rates of P-CPM and SC-CPM coded systems at 1.5
bits/s/Hz. ${\cal Y} = 64$, REC frequency pulse.}
\label{fig:C64se-15-REC}
\end{figure}

\begin{figure}[!ht]
\centering
\includegraphics[angle=270,width=3.5in]{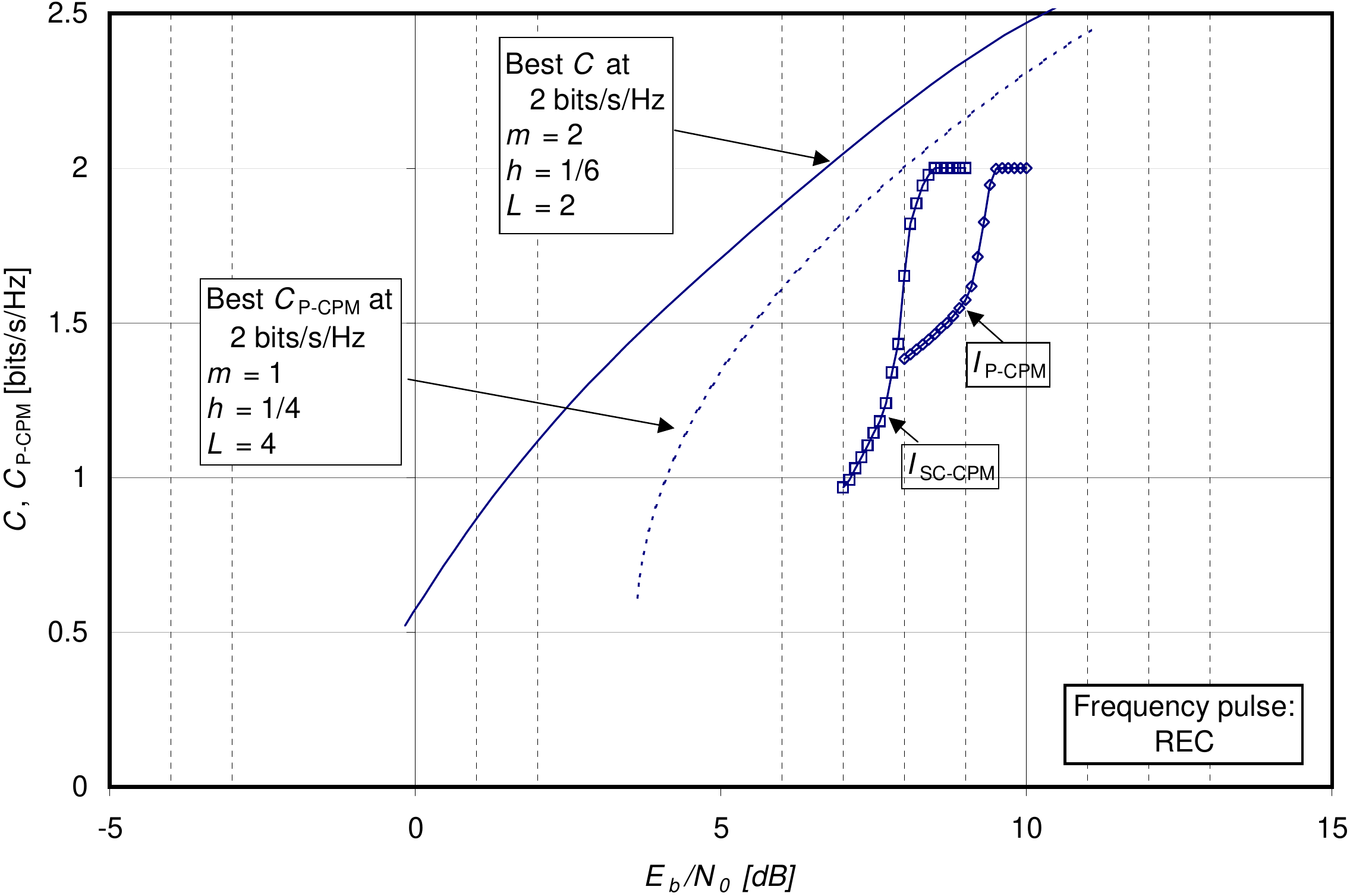}
\caption{Information rates of P-CPM and SC-CPM coded systems at 2
bits/s/Hz. ${\cal Y} = 64$, REC frequency pulse.}
\label{fig:C64se-2-REC}
\end{figure}

The information rates $I_{\rm SC-CPM}$ (resp. $I_{\rm P-CPM}$) of the SC-CPM (resp. P-CPM)
schemes has been evaluated through simulation by computing the mutual information between
the information bits at the input of the channel encoder and the soft outputs
of the channel decoder.
Fig.~\ref{fig:C64se-1-REC}, Fig.~\ref{fig:C64se-15-REC} and Fig.~\ref{fig:C64se-2-REC}
show the obtained information rates for REC frequency pulse and complexity
${\cal Y} \leq 64$.   We observe that, for a capacity of 1 bits/s/Hz,
the pragmatic scheme achieves the desired information rate
at a lower $E_b/N_0$ than the SC-CPM scheme, and at 1 dB from the pragmatic capacity.
For capacities of 1.5 bits/s/Hz and 2 bits/s/Hz, the SC-CPM scheme
achieves the desired information rate at a lower $E_b/N_0$ than the
P-CPM scheme.

\begin{figure}[!ht]
\centering
\includegraphics[angle=270,width=3.5in]{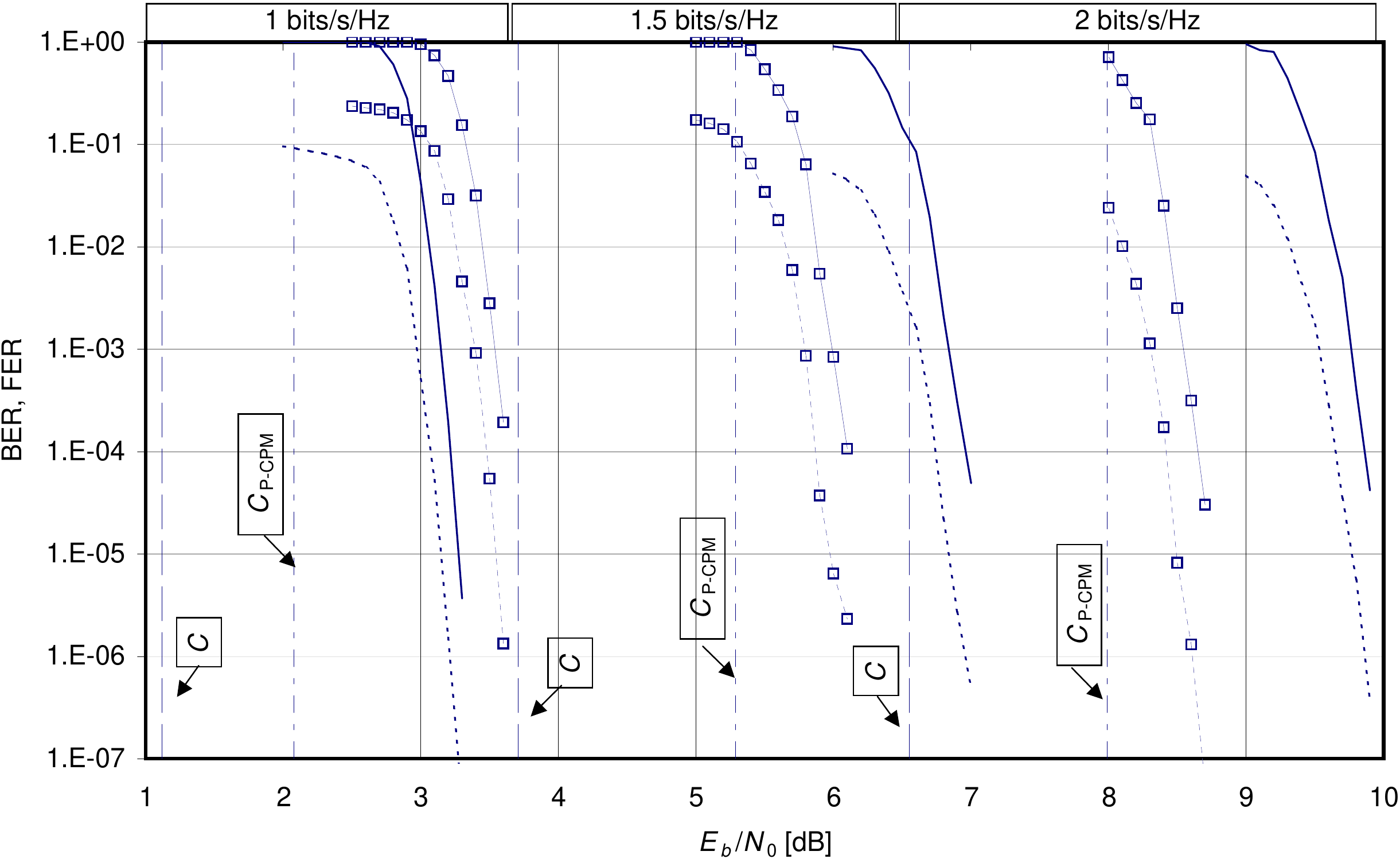}
\caption{Error rates of P-CPM (curves without markers) and SC-CPM
(curves with markers) coded systems. ${\cal Y} = 64$, REC frequency
pulse.}
\label{fig:C64er-REC}
\end{figure}

The best SC-CPM and P-CPM for the chosen spectral efficiencies have been simulated over
an AWGN channel so as to obtain accurate estimates of the bit and frame error
probabilities at very low values. The results are shown in Fig.~\ref{fig:C64er-REC}.
We observe that, at 1 bits/s/Hz, the P-CPM scheme, while featuring a lower complexity
(see Tab.~\ref{tab:complexity}), exhibits a 0.5 dB gain over the SC-CPM scheme at
FER = $10^{-4}$.   At higher spectral efficiencies, the SC-CPM scheme exhibits a gain of more than
1 dB over the P-CPM scheme.

\begin{figure}[!ht]
\centering
\includegraphics[angle=270,width=3.5in]{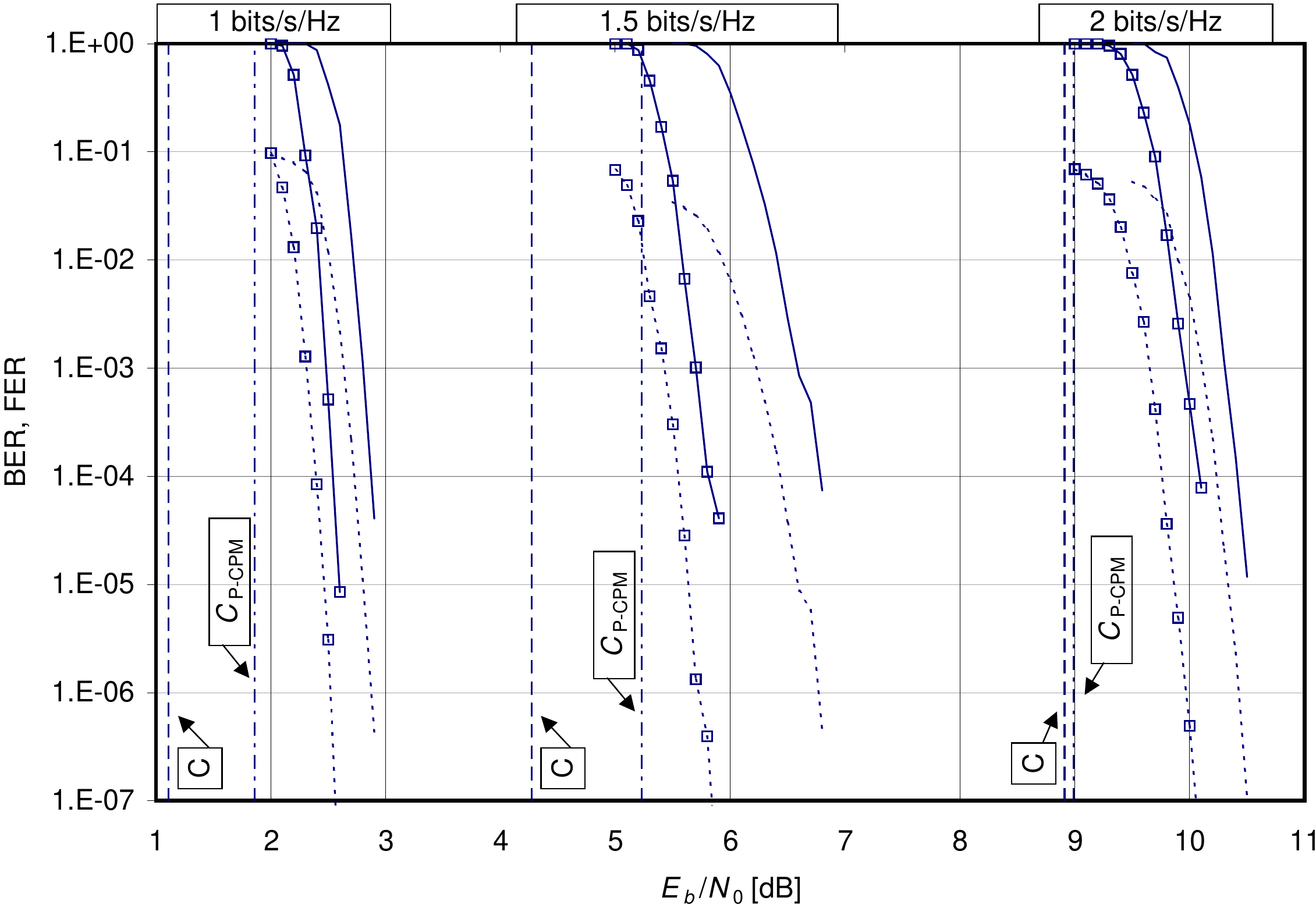}
\caption{Error rates of P-CPM (curves without markers) and SC-CPM
(curves with markers) coded systems. ${\cal Y} = 64$, RC frequency
pulse.}
\label{fig:C64er-RC}
\end{figure}

Fig.~\ref{fig:C64er-RC} shows similar results for RC frequency pulses.
The SC-CPM schemes feature a 0.3 dB gain over the P-CPM schemes at 1 bits/s/Hz and
2 bits/s/Hz, while at 1.5 bits/s/Hz the gain of the SC-CPM scheme is about 1 dB.
In both frequency pulses and all spectral efficiencies, no error floors have been observed.

\begin{figure}[!ht]
\centering
\includegraphics[angle=270,width=3.5in]{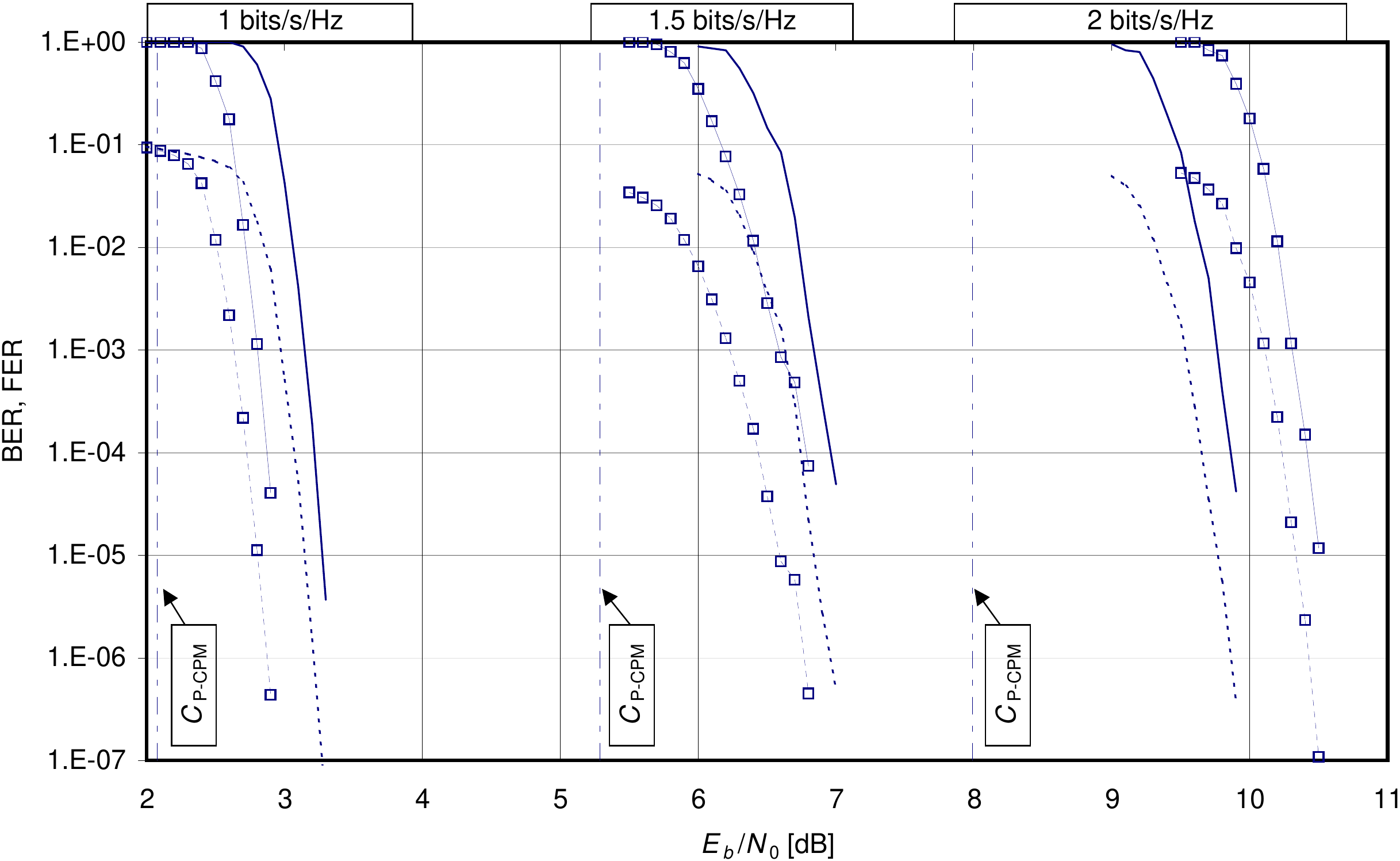}
\caption{Comparison of REC (curves without markers) and RC (curves
with markers) P-CPM schemes with $\mathcal{Y} \leq 64$.}
\label{fig:C64er-RECvsRC}
\end{figure}

A comparison of REC and RC pragmatic schemes is shown in Fig.~\ref{fig:C64er-RECvsRC}.
We observe that the RC scheme performs better than REC scheme at 1 bits/s/Hz and 1.5 bits/s/Hz.
At 2 bits/s/Hz, the hierarchy is reversed.
At 2 bits/s/Hz, REC schemes feature a gain of roughly 0.6 dB over the RC schemes at FER = $10^{-4}$.

In order to perform a fair comparison, the decoding complexity must be
kept into account.
Using definitions (\ref{eq:complexity-pcpm}) and (\ref{eq:complexity-sccpm}), the
decoding complexities of the two schemes have been computed.\
For the P-CPM scheme, the complexity of the MHOMS~\cite{bib:MHOMS} binary
decoder has been computed as follows:

\begin{equation}
\mathcal{Y}_{\rm SCCC} = 2 N_{it}\left(N_{so} + \frac{3}{2}
N_{si}\right)
\label{eq:SCCCComplexity}
\end{equation}

\noindent where the factor $3/2$ derives from the outer code rate of
the MHOMS scheme, which is 2/3.   Results are shown in
Tab.~\ref{tab:complexity}: we
observe that the complexity of the SC-CPM scheme is always larger than
that of the P-CPM scheme.   We define the complexity ratio $R_{\cal Y}
= \mathcal{Y}_{\rm SC-CPM} / \mathcal{Y}_{\rm P-CPM}$: its value is larger
than 2 and exceeds 3 for a 1 bits/s/Hz capacity with REC frequency pulse.
Higher coding gains could be achieved increasing the P-CPM complexity,
thus reducing the performance gap with respect to SC-CPM schemes.

\begin{table}[!ht]
\centering
\begin{tabular}{r|c||cc|c|c|c}
& \multirow{2}{*}{Capacity}& \multicolumn{3}{c|}{$P-CPM$} & $SC-CPM$ & Ratio\\
&  & $\mathcal{Y}_{SCCC}$ & $\mathcal{Y}_{CPM}$ &
$\mathcal{Y}_{P-CPM}$ & $\mathcal{Y}_{SC-CPM}$ & $R_{\cal Y}$\\
\hline \hline
\multirow{3}{*}{\begin{sideways}REC\end{sideways}} & 1   &  200  &  21  & 221 &  694 & 3.14  \\
& 1.5 &  200  &  41  & 241 &  709 & 2.94 \\
& 2   &  200  &  87  & 287 &  728 & 2.54 \\
\hline
\multirow{3}{*}{\begin{sideways}RC\end{sideways}} & 1   &  200  &  51  & 251 &  592 & 2.36  \\
& 1.5 &  200  &  34  & 234 &  565 & 2.41 \\
& 2   &  200  &  76  & 276 &  733 & 2.66
\end{tabular}
\caption{Complexity of P-CPM and SC-CPM schemes. ${\cal Y} = 64$.}
\label{tab:complexity}
\end{table}

%%%%%%%%%%%%%%%%%%%%%%%%%%%%%%%%%%%%%%%%%%%%%%%%%%%%%%%%%%%%%%%%%%%%%%
\section{Conclusions}
\label{sec:conclusions}
%%%%%%%%%%%%%%%%%%%%%%%%%%%%%%%%%%%%%%%%%%%%%%%%%%%%%%%%%%%%%%%%%%%%%%

The pragmatic approach to coded continuous-phase modulation (CPM) has been proposed
as a capacity-approaching low-complexity alternative to the serially-concatenated
CPM (SC-CPM) coding scheme.
After performing a selection of the best spectrally-efficient
CPM modulations to be embedded into SC-CPM schemes, the pragmatic
capacity  of CPM modulations has been evaluated and  optimized  through a careful
design of the mapping between input bits and CPM waveforms.
The so obtained schemes have been cascaded with an outer serially-concatenated convolutional
code to form a pragmatic coded-modulation system.
The resulting schemes have been shown to exhibit performance  close to the CPM capacity without
requiring iterations between the outer decoder and the CPM demodulator. As a result,
the receiver exhibits reduced complexity and increased flexibility due to the separation
of the demodulation and decoding functions.

%%%%%%%%%%%%%%%%%%%%%%%%%%
\section*{Acknowledgements}
%%%%%%%%%%%%%%%%%%%%%%%%%%

This work has been supported by Regione Piemonte under Contract E4.

%\nocite{*}
%-----------------------------------------------------------------
%GATHER{../../bib/Xbib.bib}   % For Gather Purpose Only
%GATHER{../../bib/Xbib.bbl}   % For Gather Purpose Only
\bibliographystyle{IEEEtran}
% \bibliography{IEEEabrv,../../bib/xbib}

\begin{appendices}

\section{Proof of Theorem \ref{th:1}}\label{sec:App1}

First, we revise some properties of the CPM trellis, for a
\emph{binary} CPM scheme.

An error event generated  by a given difference sequence
$\mathbf{b}=(b_1,\dots,b_{\Delta})\in \{-1,0,1\}^{\Delta}$, is a
pair of trellis paths with length $\Delta + L -1$ trellis steps,
which diverge at, say, time $0$ and merge back together at time
$\Delta + L -1$. For a given value of $\mathbf{b}$, there exist
several different error events, corresponding to all possible
choices of the variables:
\begin{equation}\label{eq:a1}
\beta_0, a_{-L+2}, \dots, a_0, a_1^{(1)}, \dots, a_{\Delta}^{(1)},
a_{\Delta +1}, \dots, a_{\Delta+L-1}
\end{equation}
and
\begin{equation}\label{eq:a2}
\beta_0, a_{-L+2}, \dots, a_0, a_1^{(2)}, \dots, a_{\Delta}^{(2)},
a_{\Delta +1}, \dots, a_{\Delta+L-1}
\end{equation}
where $\beta_0, a_{-L+2}, \dots, a_0$ determine the starting
state,
\begin{equation}
a_n^{(1)} = \left\{
\begin{array}{cc}
1 & b_n = 1 \\
0 & b_n = -1 \\
a_n & b_n = 0
\end{array} \right.
\end{equation}
and
\begin{equation}
a_n^{(2)} = \left\{
\begin{array}{cc}
0 & b_n = 1 \\
1 & b_n = -1 \\
a_n & b_n = 0
\end{array} \right.,
\end{equation}
and $a_{\Delta +1}, \dots, a_{\Delta+L-1}$ determine the final
state of the error event. If two trellis edges belong to the same
trellis section of the error event, we call them an \emph{edge
pair} of the error event.

Now, we build the graph $\mathcal{G}(\mathbf{b})=(\mathcal{V},
\mathcal{E})$ as described in Sect. \ref{sec:pragcap}: two
vertices are adjacent if and only if the corresponding trellis
edges do not have the same starting state and constitute an edge
pair of an error event generated by $\mathbf{b}$. The graph
$\mathcal{G}(\mathbf{b})$ has
 $\mathcal{N}(\mathbf{b})$ connected components, denoted
$\mathcal{C}_0, \dots, \mathcal{C}_{\mathcal{N}(\mathbf{b})-1}$,
where $\mathcal{C}_i=(\mathcal{V}_i, \mathcal{E}_i)$, whose
properties will be investigated hereafter.

For an integer amount $\gamma$, the $\gamma$-\emph{rotation}
transforms the trellis edge $(\mathbf{\alpha}, \beta)$ into the
trellis edge $(\mathbf{\alpha}, (\beta + \gamma)_P)$. The image of
a set of edges through the $\gamma$-rotation is the set of images
of the edges belonging to the set through the $\gamma$-rotation.
The following proposition holds:
\begin{proposition} \label{prop:rot}
For any $\gamma$, the $\gamma$-rotation transforms the set
$\mathcal{V}_i$, corresponding to a component $\mathcal{C}_i$ of
$\mathcal{G}(\mathbf{b})$, into a set $\mathcal{V}_j$,
corresponding to another component $\mathcal{C}_j$ (even,
possibly, the same).
\end{proposition}
\proof Two trellis edges $(\mathbf{\alpha}, \beta)$ and
$(\mathbf{\alpha}', \beta')$ are in the same component
$\mathcal{C}_i$ if and only if in the graph there is a
length-$\kappa$ path,
\[
(\mathbf{\alpha}^{(1)}, \beta^{(1)}) \rightarrow
(\mathbf{\alpha}^{(2)}, \beta^{(2)}) \rightarrow \dots
 \rightarrow (\mathbf{\alpha}^{(\kappa)},
\beta^{(\kappa)})
\]
with $(\mathbf{\alpha}^{(1)}, \beta^{(1)}) = (\mathbf{\alpha},
\beta)$ and $(\mathbf{\alpha}^{(\kappa)}, \beta^{(\kappa)}) =
(\mathbf{\alpha}', \beta')$. This means that, for
$i=1,\dots,\kappa-1$, $(\mathbf{\alpha}^{(i)}, \beta^{(i)})$ and
$(\mathbf{\alpha}^{(i+1)}, \beta^{(i+1)})$ have different starting
states and form an edge pair of an error event generated by
$\mathbf{b}$.

Now, notice that, given an error event generated by $\mathbf{b}$,
by changing $\beta_0$ in \refeq{a1} and \refeq{a2} into $(\beta_0
+ \gamma)_P$, we obtain another error event generated by
$\mathbf{b}$, whose edge pairs are in one-to-one correspondence
through $\gamma$-rotation with the edge pairs of the original
error event. For this reason, in the graph there is the path
\[
(\mathbf{\alpha}^{(1)}, \beta^{(1)}_{\gamma}) \rightarrow
(\mathbf{\alpha}^{(2)}, \beta^{(2)}_{\gamma}) \rightarrow \dots
 \rightarrow (\mathbf{\alpha}^{(\kappa)},
\beta^{(\kappa)}_{\gamma})
\]
where $\beta^{(i)}_{\gamma} \triangleq (\beta^{(i)} + \gamma)_P$.

Thus, by definition, $(\mathbf{\alpha}, (\beta+\gamma)_P)$ and
$(\mathbf{\alpha}', (\beta'+\gamma)_P)$ belong to the same
component $\mathcal{C}_j$. For the arbitrariness of
$(\mathbf{\alpha}, \beta)$ and $(\mathbf{\alpha}', \beta')$, we
have that the image of $\mathcal{V}_i$ through $\gamma$-rotation
is contained in $\mathcal{V}_j$.

Finally, notice that, with the same argument, the image of
$\mathcal{V}_j$ through $(P-\gamma)$-rotation is contained in
$\mathcal{V}_i$. From this fact, and the injectivity of rotations,
we deduce that the image of $\mathcal{V}_i$ through
$\gamma$-rotation is equal to $\mathcal{V}_j$.
\endproof

From \refeq{a1} and \refeq{a2}, we can deduce a set of
\emph{rules} for a pair of trellis edges to be in the same
component, in the following way:
\begin{itemize}
\item We fix a trellis section inside the error event. The error
event has length $\Delta+L-1$ trellis sections, but the first is
not to be considered, thus we have $\Delta+L-2$ possible choices.

\item We consider the constraints imposed on an edge pair of the
error event at that trellis section.
\end{itemize}

We derive two types of rules, according to the trellis section we
are considering.

{\bf Type-I rules:} They correspond to the last $L$ trellis
sections of the error event. They give the following edge pairs,
for $j = \Delta-L+1, \dots, \Delta$:
\begin{equation} \label{eq:a1bis}
\left(\mathbf{\alpha}_j^{(1)}, \left(\beta_{\Delta+L-1} -Q
w_H\left(\mathbf{\alpha}_j^{(1)}\right)\right)_P\right)
\end{equation}
and
\begin{equation} \label{eq:a2bis}
\left(\mathbf{\alpha}_j^{(2)}, \left(\beta_{\Delta+L-1} -Q
w_H\left(\mathbf{\alpha}_j^{(2)}\right)\right)_P\right),
\end{equation}
where $\mathbf{\alpha}_j^{(i)}=(a_j^{(i)},\dots,a_{j+L-1}^{(i)})$,
$a_j^{(i)}=a_j$ if $j<1$ or $j > \Delta$, and
\[
\beta_{\Delta+L-1} = \left(\beta_0 + Q\sum_{i=-L+2}^{\Delta}
a_i^{(1)}\right)_P.
\]
The above relation holds because $\sum_{i=1}^{\Delta} a_i^{(1)} =
\sum_{i=1}^{\Delta} a_i^{(2)} \mod P$.

{\bf Type-II rules:} They correspond to the $\Delta-2$ trellis
sections of the error event from the second to the $(\Delta-1)$-th
one. They give the following edge pairs, for $j = -L+3, \dots,
\Delta-L$:
\begin{equation} \label{eq:a1ter}
\left(\mathbf{\alpha}_j^{(1)}, \beta_{j+L-1}^{(1)}\right)
\end{equation}
and
\begin{equation} \label{eq:a2ter}
\left(\mathbf{\alpha}_j^{(2)}, \beta_{j+L-1}^{(2)}\right),
\end{equation}
where $\mathbf{\alpha}_j^{(i)}=(a_j^{(i)},\dots,a_{j+L-1}^{(i)})$,
$a_j^{(i)}=a_j$ if $j<1$ or $j > \Delta$, and
\[
\beta_{j+L-1}^{(i)} = \left(\beta_0 + Q\sum_{k=-L+2}^{j}
a_k^{(i)}\right)_P.
\]
Notice that, if $\Delta = 2$, there are no type-II rules.

The following lemma will be the key to prove Theorem \ref{th:1}.
\begin{lemma} \label{lemma:1}
Consider the sets of trellis edges:
\begin{equation}
\widetilde{\mathcal{V}}_i = \left\{\left(\mathbf{\alpha}, (i- Q
w_H(\mathbf{\alpha}))_P \right) : \mathbf{\alpha} \in \{0,1\}^L
\right\},
\end{equation}
for $i=0,\dots,P-1$. Each of these sets is entirely contained in
one connected component of the graph $\mathcal{G}(\mathbf{b})$.
\end{lemma}
\proof We prove by induction that the set of type-I rules implies
the lemma, starting from the last rule (i.e., $j=\Delta$ in
\refeq{a1bis} and \refeq{a2bis}) and going backwards.

Suppose that the last $d$ rules ($d<L$) imply that the set of
edges of the form:
\[
\widetilde{\mathcal{V}}_{\beta_{\Delta+L-1},d}(\mathbf{x}) =
\left\{\left([\mathbf{d},\mathbf{x}], \left(\beta_{\Delta+L-1} -Q
w_H([\mathbf{d},\mathbf{x}])\right)_P\right), \mathbf{d} \in
\{0,1\}^d\right\}
\]
for a fixed length-$(L-d)$ vector
$\mathbf{x}=(x_1,\dots,x_{L-d})$, is entirely contained in one
connected component. This is true for $d=1$, since $b_{\Delta}
\neq 0$.

Now, again because $b_{\Delta} \neq 0$, the last-but-$(d+1)$ rule
pairs edges like $\mathbf{d}' x_1 \dots x_{L-d}$ and $\mathbf{d}''
\overline{x_1} \dots x_{L-d}$, for suitable length-$d$ vectors
$\mathbf{d}'$ and $\mathbf{d}''$, where $\overline{x}$ is the
complement of $x$. This implies that
\[
\widetilde{\mathcal{V}}_{\beta_{\Delta+L-1},d+1}(\mathbf{x}) =
\widetilde{\mathcal{V}}_{\beta_{\Delta+L-1},d}([0,\mathbf{x}])
\cup
\widetilde{\mathcal{V}}_{\beta_{\Delta+L-1},d}([1,\mathbf{x}]),
\]
for a fixed length-$(L-d-1)$ vector $\mathbf{x}$, is entirely
contained in one connected component.

By induction up to $d=L$, we have proved that, for all possible
values of $\beta_{\Delta+L-1}$,
$\widetilde{\mathcal{V}}_{\beta_{\Delta+L-1}} =
\widetilde{\mathcal{V}}_{\beta_{\Delta+L-1},L}$ is entirely
contained in one connected component.
\endproof

\proof [Of Theorem \ref{th:1}] If $\Delta(\mathbf{b})=2$, then
there are only type-I rules, and the components of
$\mathcal{G}(\mathbf{b})$ satisfy, for $i=0,\dots,P-1$:
\begin{equation}
\mathcal{V}_i=\widetilde{\mathcal{V}}_i
\end{equation}
and the theorem is proved.

If $\Delta(\mathbf{b})>2$, instead, there are also type-II rules.
Consider, in particular, the type-II rule with $j=\Delta-L$ in
\refeq{a1ter} and \refeq{a2ter}. It gives the edge pair:
\begin{equation} \label{eq:a1ter1}
\left(\mathbf{\alpha}_{\Delta-L}^{(1)},
\beta_{\Delta-1}^{(1)}\right)
\end{equation}
and
\begin{equation} \label{eq:a2ter1}
\left(\mathbf{\alpha}_{\Delta-L}^{(2)},
\beta_{\Delta-1}^{(2)}\right).
\end{equation}
But the above transitions belong to
$\widetilde{\mathcal{V}}_{(\beta_{\Delta+L-1}-Q)_P}$ and to
$\widetilde{\mathcal{V}}_{\beta_{\Delta+L-1}}$. Then, these two
sets are merged by the type-II rule with $j=\Delta-L$. Consider
the component $\mathcal{C}_1$ of $\mathcal{G}(\mathbf{b})$ that
contains these sets and let $\mathcal{V}_1$ be its vertex set.
Prop. \ref{prop:rot} implies that a $Q$-rotation transforms
$\mathcal{V}_1$ into itself. But then, $\mathcal{C}_1$ also
contains $\widetilde{\mathcal{V}}_{(\beta_{\Delta+L-1}-lQ)_P}$,
for every integer $l$. Since $Q$ and $P$ are 3, we deduce
that $\mathcal{C}_1$ contains all $\widetilde{\mathcal{V}}_{i}$'s
and thus is the only component of $\mathcal{G}(\mathbf{b})$. This
concludes the proof.
\endproof

\section{Proof of Theorem \ref{th:2}} \label{sec:App2}
To prove Theorem \ref{th:2}, we first notice that Prop.
\ref{prop:rot} holds for a 1 CPM scheme also, since in the
proof, the hypothesis of binary input symbols is not used.

Also, there are two types of rules, which are the extension to the
1 case of the two types introduced in the previous
sections.

{\bf Type-I rules:} They correspond to the last $L$ trellis
sections of the error event. They give the following edge pairs,
for $j = \Delta-L+1, \dots, \Delta$:
\begin{equation} \label{eq:a1bis1}
\left(\mathbf{\alpha}_j^{(1)}, \left(\beta_{\Delta+L-1} -Q
\sum_{i=0}^{L-1}\alpha_{j+i}^{(1)}\right)_P\right)
\end{equation}
and
\begin{equation} \label{eq:a2bis1}
\left(\mathbf{\alpha}_j^{(2)}, \left(\beta_{\Delta+L-1} -Q
\sum_{i=0}^{L-1}\alpha_{j+i}^{(2)}\right)_P\right),
\end{equation}
where $\mathbf{\alpha}_j^{(i)}=(a_j^{(i)},\dots,a_{j+L-1}^{(i)})$,
$a_j^{(i)}=a_j$ if $j<1$ or $j > \Delta$, and
\[
\beta_{\Delta+L-1} = \left(\beta_0 + Q\sum_{i=-L+2}^{\Delta}
a_i^{(1)}\right)_P.
\]
The above relation holds because $\sum_{i=1}^{\Delta} a_i^{(1)} =
\sum_{i=1}^{\Delta} a_i^{(2)} \mod P$.

{\bf Type-II rules:} They correspond to the $\Delta-2$ trellis
sections of the error event from the second to the $(\Delta-1)$-th
one. They give the following edge pairs, for $j = -L+3, \dots,
\Delta-L$:
\begin{equation} \label{eq:a1ter2}
\left(\mathbf{\alpha}_j^{(1)}, \beta_{j+L-1}^{(1)}\right)
\end{equation}
and
\begin{equation} \label{eq:a2ter2}
\left(\mathbf{\alpha}_j^{(2)}, \beta_{j+L-1}^{(2)}\right),
\end{equation}
where $\mathbf{\alpha}_j^{(i)}=(a_j^{(i)},\dots,a_{j+L-1}^{(i)})$,
$a_j^{(i)}=a_j$ if $j<1$ or $j > \Delta$, and
\[
\beta_{j+L-1}^{(i)} = \left(\beta_0 + Q\sum_{k=-L+2}^{j}
a_k^{(i)}\right)_P.
\]
Notice that, if $\Delta = 2$, there are no type-II rules.

The following lemma is the extension of Lemma \ref{lemma:1} to the
1 case.

\begin{lemma}
Let $\mathbf{b}$ be a difference sequence of length $\Delta$ with
$b_{\Delta} = \pm 1$. Consider the sets:
\begin{equation}
\widetilde{\mathcal{C}}_i = \left\{\left(\mathbf{\alpha}, \left(i-
Q \sum_{i=1}^L \alpha_i\right)_P \right) : \mathbf{\alpha} \in
\{0,\dots,M-1\}^L \right\},
\end{equation}
for $i=0,\dots,P-1$. Each of these sets is entirely contained in
one connected component of the graph $\mathcal{G}(\mathbf{b})$.
\end{lemma}
\proof By induction on the set of type-I rules. It is the
straightforward extension of Lemma \ref{lemma:1}.
\endproof

Finally, the proof of Theorem \ref{th:2} follows the proof of
Theorem \ref{th:1}, by noticing that if $\Delta(\mathbf{b})=2$,
then there are only type-I rules, while if $\Delta(\mathbf{b})>2$
there are also type-II rules, that cause the graph
$\mathcal{G}(\mathbf{b})$ to be connected.

\end{appendices}

\end{document}